\documentclass[11pt,a4paper]{article}

\usepackage[T1]{fontenc}
\usepackage[utf8]{inputenc}
\usepackage{lmodern}

\usepackage{jheppub}

\usepackage{graphicx,amssymb,amsmath,amsfonts}
\usepackage{amsthm}

\usepackage{hyperref}

\usepackage{subcaption}
\usepackage{xcolor}
\usepackage{mathrsfs}
\usepackage{slashed} 

\newcommand{\ab}{{\alpha\beta}}
\let\a\relax
\let\b\relax
\let\c\relax
\let\d\relax
\newcommand{\a}{\alpha}
\newcommand{\b}{\beta}
\newcommand{\c}{\gamma}
\newcommand{\d}{\delta}

\newcommand{\abcd}{{\alpha\beta\gamma\delta}}

\newcommand{\V}{\mathbf}

\newcommand{\p}{{\partial}}
\newcommand{\w}{{\omega}}

\newcommand{\ep}{\epsilon}

\newcommand{\zb}{{\bar{z}}}

\newcommand{\gzz}{{\gamma_{z\bar{z}}}}
\newcommand{\guzz}{{\gamma^{z\bar{z}}}}

\newcommand{\mH}{\mathcal{H}}
\newcommand{\mI}{\mathcal{I}}

\newcommand{\mL}{\mathcal{L}}

\newcommand{\mN}{\mathcal{N}}

\newcommand{\mQ}{\mathcal{Q}}

\newcommand{\mW}{\mathcal{W}}

\let\O\relax
\newcommand{\O}{O}

\let\t\relax
\newcommand{\t}{\theta}

\newcommand{\cc}{\text{c.c.}}

\let\tilde\relax
\newcommand{\tilde}{\widetilde}

\DeclareMathOperator{\tr}{tr}
\DeclareMathOperator{\diag}{diag}

\let\Re\relax
\let\Im\relax
\DeclareMathOperator{\Re}{Re}
\DeclareMathOperator{\Im}{Im}
\newcommand{\C}{{\mathbb{C}}}

\newcommand{\rhs}{{r$.$h$.$s$.$}}

\newcommand{\s}{{\mH^+_-}}
\newcommand{\pS}{{\p \Sigma}}
\let\H\relax
\newcommand{\H}{{\mH^+}}

\newcommand{\QEfull}[1][f]{{Q_{#1}^\H}}
\newcommand{\QMfull}[1][f]{{\tilde Q_{#1}^\H}}
\newcommand{\QE}[1][f]{{\slashed\delta\QEfull[#1]}}
\newcommand{\QM}[1][f]{{\slashed\delta\QMfull[#1]}}
\newcommand{\iQE}[1][f]{{\delta\QEfull[#1]}}
\newcommand{\iQM}[1][f]{{\delta\QMfull[#1]}}
\newcommand{\nQE}[1][f]{{\mN_{#1}^\H}}

\let\l\relax
\newcommand{\l}{\lambda}
\newcommand{\eQE}[1][\l]{{\mQ_{#1}^\H}}
\newcommand{\eQM}[1][\l]{{\tilde\mQ_{#1}^\H}}

\newcommand{\GF}{{\Delta}}

\newcommand{\x}{\Theta}

\title{Singular Supertranslations and Chern-Simons Theory on the Black Hole Horizon}

\author[a]{Ratindranath Akhoury,}
\author[b]{Sangmin Choi}
\author[c,d,e]{and Malcolm J. Perry}

\affiliation[a]{Leinweber Center for Theoretical Physics, Randall Laboratory of Physics,\\ Department of Physics, University of Michigan, Ann Arbor, MI 48109, USA.}
\affiliation[b]{CPHT, CNRS, \'Ecole Polytechnique,\\ Institut Polytechnique de Paris, F-91128 Palaiseau, France.}
\affiliation[c]{School of Physics and Astronomy, Queen Mary University of London,\\ Mile End Road, London E1 4NS, UK.}
\affiliation[d]{DAMTP, Centre for Mathematical Sciences, Cambridge University,\\ Wilberforce Road, Cambridge CB3 OWA, UK.}
\affiliation[e]{Trinity College, Cambridge, CB2 1TQ, UK.}

\emailAdd{akhoury@umich.edu}
\emailAdd{sangmin.choi@polytechnique.edu}
\emailAdd{malcolm@damtp.cam.ac.uk}

\abstract{
We construct the standard and dual supertranslation charges on the future horizon of the Schwarzschild black hole,
	using the first-order formulation of gravity with the Holst action.
The Dirac bracket algebra of standard and dual supertranslation charges is shown to exhibit a central term in the presence of singularities
	in the two-sphere function associated with supertranslation.
We show that one can cancel this anomalous term and restore the asymptotic symmetry algebra by
	introducing a gravitational Chern-Simons theory on the horizon.
This demonstrates that consistency of the asymptotic symmetry algebra requires a new structure on the horizon.
}

\begin{document}

{
	\maketitle
}

\section{Introduction}

Black hole physics provides us with a paradox that has been around now for almost fifty years \cite{Hawking:1976ra}.
The information paradox illustrates an apparent conflict between classical and semi-classical 
general relativity and the fundamental tenets of quantum theory. The classical black hole uniqueness 
theorems appear to indicate that the Kerr-Newman family of black holes, characterised by their
mass, angular momentum and electric charge, is sufficient to describe all black hole stationary 
states\footnote{There are some exceptions to this, but they do not change the general picture.}.
If this were true in a complete quantum theory, then it would not be possible to 
distinguish between a black hole formed from matter and one formed form antimatter. It appears
that to any observer outside a black hole that has reached a stationary state, the black hole is 
independent of the details of its formation. In particular, the black hole has no memory of the
quantum state of the material that formed it.  The trouble comes when black holes evaporate. 
Hawking showed that the outgoing radiation is thermal and so has a large von Neumann entropy. Suppose
that the matter forming the black hole was in a pure quantum state. In  quantum mechanics, the von
Neumann entropy is constant because the time evolution operator is unitary but that is inconsistent
with the picture outlined above. Identifying what is wrong with this picture has been a huge 
challenge and, despite much hard labour, has not yet yielded any clear solution.

Recently, it has been realised that black holes can have soft hair \cite{Hawking:2016msc,Hawking:2016sgy}. Soft hair are extra degrees of freedom
that a black hole can have. The geometry remains that of the Kerr-Newman sequence with the soft hair
being described by a particular class of gauge transformations.   Suppose we look at an 
asymptotically flat spacetime that does not contain a black hole. Bondi-Metzner-Sachs (BMS) transformations
acting on the gravitational field at both past and future null infinity generalise the Poincar\'e 
symmetry group familiar from non-gravitational settings \cite{Bondi:1962px,Sachs:1962wk}. The Poincar\'e group acts on Minkowski
space with large gauge transformations generating translations, rotations or boosts. Each of these
large gauge transformations is associated with a charge namely the momentum, angular momentum and the
boost charge. Similarly, the BMS transformations are associated with a charge conjugate to 
large gauge transformations. These charges distinguish the infinite number of distinct vacua of 
the gravitational field.  In electromagnetism, one is familiar with integrating a current over a 
spacelike three-surface to describe the charge passing through the surface. Gauss' theorem then 
guarantees that this integral can be turned into a surface integral that measures the charge inside 
that surface. Exactly the same thing happens for the BMS charges so they can be described by 
surface integrals on sections of past or future null infinity. These charges can change 
as the result of incoming matter or gravitational waves passing through past null infinity
or outgoing matter or gravitational waves passing through future null infinity. Gravitational 
memory gives a method of observing the changes in these charges.
We refer interested readers to
	\cite{Strominger:2013lka,Strominger:2013jfa,He:2014laa,Hyun:2014kfa,Adamo:2014yya,He:2014cra,Campiglia:2015kxa,
	Campiglia:2015lxa,Campiglia:2015qka,Campiglia:2015yka,Kapec:2015ena,Avery:2015gxa,Avery:2015iix,Lysov:2015jrs}
	for some earlier literature on this development.
See \cite{Strominger:2017zoo} for a review.

In black hole spacetimes, the horizon is a boundary of what can be observed form the
exterior.  The integrals of currents
may then have two boundary components, one at null infinity and the other on the horizon. 
As a consequence black holes will also carry soft charges in much the same way as they can be found at
null infinity. This paper is concerned with some of the consequences of this observation.  
Our main aim here is to consider pure gravity without matter.
However, in the interests of clarity 
and simplicity, we will also provide an outline discussion of the case of electromagnetism
as a model of the more complicated case of gravitation.
The main results of this paper have been outline in the letter \cite{Akhoury:2022sfj}, and in this paper we provide details of the calculation.

We examine in detail the physics of horizon (standard and dual) BMS charges for the Schwarzschild black hole.
The horizon charges are computed using the machinery presented by \cite{Godazgar:2020gqd,Godazgar:2020kqd} in the first-order formalism of gravity.
The algebra of the charges is expected
to reflect the algebra of the vector fields that generate the corresponding symmetry. We find 
an anomaly in the algebra of charges. To preserve the symmetry of the theory, we need to introduce 
some degrees of freedom to cancel the anomaly since otherwise the theory would be inconsistent. 
We show that this can be done by the introduction of a (holographic) gravitational Chern-Simons theory on the horizon.
It would be satisfying to show that the states of this Chern-Simons theory reproduce the 
correct black hole entropy and thereby describe the states of the black hole itself. Such a goal
is currently beyond us but a subject of current investigation\footnote{Were this true in the most 
obvious simple way, it would appear run into difficulties because of the species problem. The Chern-Simons 
theory on the boundary depends not only on the gravitational degrees of freedom but also on the matter
degrees of freedom. So you would expect the spectrum of states to depend on the entropy. However,
the black hole entropy as given by Hawking, is just one quarter of the area of the event horizon 
and does not depend on the matter content. All state counting arguments appear to run into 
this type of difficulty.}. We find the Chern-Simons theory for electromagnetism to have gauge group
 $U(1)\otimes U(1)$ and for gravitation to be $SL(2,\C)$ \cite{Witten:1989ip}.
Thus from this point of view, consistency requires the introduction of a new structure at the horizon.
 
 In section \ref{sec:BMS}, we describe the analog of standard and dual BMS transformations on the horizon in the Bondi gauge.
 The use of the Bondi gauge on the horizon makes computations particularly simple for the case
 of the Schwarzschild metric. It is noteworthy that the algebra of vector fields that 
 generate supertranslations and superrotations is identical to that found for the BMS group at
 null infinity. However, to establish this result, we had to revisit some earlier work of 
 Barnich and Troessaert where a modified Lie bracket was introduced; the rationale and description
 is also discussed in section \ref{sec:BMS}.  In section \ref{sec:charge}, we introduce the charges associated to 
 the diffeomorphism symmetries. In parallel, we also discuss the dual (magnetic) counterpart of the
 diffeomorphism symmetries. We restrict ourselves here to use of smooth vector fields to generate the
 symmetries. In section \ref{sec:Qpole}, we continue the discussion of charges but allow for the possibility
 that there could be singularities in the supertranslations. We examine in detail the case of 
 the supertranslation generator having a pole when expressed in the usual complex coordinates on 
 the $S^2$ of the horizon. In section \ref{sec:bracket}, we show that the algebra of electric and 
 magnetic supertranslation charges is anomalous and discover the nature of a central charge.
 In section \ref{sec:bracket2}, we give an alternative derivation of the same result. In section \ref{sec:em},
 we examine electromagnetic soft hair and show that a singularity lead to an anomaly in the charge algebra
 when one has both electric and magnetic transformations. We show that this anomaly can be canceled by 
 supposing that the horizon has a Chern-Simons theory living on it. It is fortunate
 that the Chern-Simons is a topological theory as it is metric independent. There
 are two nice properties that follow. The first is that since the horizon is a null surface,
 the metric is degenerate there and one cannot invert the metric. Had the theory been metric-dependent,
 as most are, it would have been impossible to formulate a theory that is restricted to the
 null surface. The second also follows from being metric-independent. The energy-momentum tensor
 of a theory is given by varying the action with respect to the metric. Therefore, in the 
 Chern-Simons case, the energy-momentum tensor vanishes and the holographic theory does not
 disturb the black hole geometry. In section \ref{sec:cs}, we repeat this analysis for the gravitational case. 
 Finally, there is a brief discussion of our results in section \ref{sec:discussion}. In addition, there are three appendices
 that deal with some technical matters involved in our computations\footnote{
 \textbf{Notation}:
	We work in units where $G_N=1$.
	We will use lower-case Latin letters $a,b,c,\ldots$ for the four-dimensional curved indices,
	Greek letters $\a,\b,\c,\ldots$ for the four-dimensional flat (Lorentz) indices,
	and capital Latin letters $A, B, C, \ldots$ for the two-dimensional curved indices corresponding to angular variables on a sphere.
	The indices $a,b,c,\ldots$ are lowered/raised by $g_{ab}$ and its inverse $g^{ab}$,
		while $\a,\b,\c,\ldots$ are lowered/raised by $\eta_\ab$ and its inverse $\eta^\ab$.
	The two-dimensional indices $A,B,C,\cdots$ are lowered and raised by the unit 2-sphere metric $\gamma_{AB}$ and its inverse $\gamma^{AB}$.
		An exception to this convention is used in appendix \ref{app:charges} 
		where $g_{AB}$ and its inverse $g^{AB}$ are used to lower and raise indices.}.


\section{Horizon BMS transformations in the Bondi gauge}\label{sec:BMS}

We briefly review the BMS supertranslations and superrotations on the future horizon of a Schwarzschild black hole \cite{Hawking:2016sgy}.
Throughout our paper, we work in the Bondi gauge,
\begin{align}
	g_{rr} = g_{rA} = 0
	,\qquad
	\p_r\det\left(\frac{g_{AB}}{r^2}\right) = 0
	.
\end{align}
In terms of the ingoing Eddington-Finkelstein coordinates, the Schwarzschild metric is given by
\begin{align}
	ds^2 = -\Lambda dv^2 + 2dvdr + r^2 \gamma_{AB} d\x^A d\x^B
	,\qquad
	\Lambda \equiv 1-\frac{2M}{r}
	,
\end{align}
where $\gamma_{AB}$ is the metric on the unit 2-sphere.
A diffeomorphism $\xi$ that preserves these conditions should satisfy
\begin{align}
	\mL_\xi g_{rr} = \mL_\xi g_{rA} = 0
	,\qquad
	\gamma^{AB}\mL_\xi g_{AB} = 0
	.
	\label{Bondi}
\end{align}
Such diffeomorphisms can be parametrized as \cite{Hawking:2016sgy}
\begin{align}
	\xi = X \p_v - \frac{1}{2}\left(rD_A X^A + D^2 X\right) \p_r + \left(X^A + \frac{1}{r}D^A X\right)\p_A
	,
	\label{general}
\end{align}
where $X^A=X^A(v,\x)$ is an arbitrary vector field and $X=X(v,\x)$ is an arbitrary scalar 
field on the future horizon $\mH^+$.
Here $D_A$ denotes the covariant derivative on the unit 2-sphere and so $D^A = \gamma^{AB}D_B$ 
and also $D^2\equiv D^AD_A = \gamma^{AB}D_AD_B$.

A supertranslation is given by
\begin{align}
	X = f(\x)
	,\qquad
	X^A = 0
	,
\end{align}
where $f$ is a smooth function on the 2-sphere.
In later sections, we relax the smoothness condition to allow $f$ to have poles.

A superrotation is given by
\begin{align}
	X = \frac{v}{2}D_A Y^A
	,\qquad
	X^A = Y^A(\x)
	,
\end{align}
where $Y^A$ is a smooth vector field on the 2-sphere.

Since supertranslations and superrotations are metric-dependent, the diffeomorphisms 
\eqref{general} do not form a closed algebra under the Lie bracket of vector fields. To see why, 
consider a  transformation of the 
metric generated by $\xi_1^a$. Under such a transformation $g_{ab} \rightarrow g_{ab}+h_{ab}$ with
\begin{align}
h_{ab} = {\mathcal L}_{\xi_1} g_{ab} = \xi_1^c\partial_cg_{ab} + g_{ac}\partial_b\xi_1^c + g_{cb}\partial_a\xi_1^c.
\end{align}
Now a second transformation generated by $\xi_2^a$ will produce the second order variation of the metric
but will also produce a variation of $\xi_1^a$. The variation 
of $\xi_1^a$ needs to be removed  in order to isolate the second order variation of the metric. 
The Lie bracket $[\xi_1,\xi_2]$ of two vector fields $\xi_1^a$ and $\xi_2^a$ is conventionally defined by
\begin{align}
 \mathcal L_{\xi_1}\mathcal L_{\xi_2} - \mathcal L_{\xi_2}\mathcal L_{\xi_1} =\mathcal  L_{[\xi_1,\xi_2]}
 \end{align}
so that
\begin{align} 
 	[\xi_1,\xi_2]^a = \xi_1^b\p_b \xi_2^a - \xi_2^b\p_b \xi_1^a.
\end{align}
The Lie bracket needs to be modified in order to isolate just the second order variation of the metric. 
An appropriately modified Lie bracket of vector fields
was 
introduced by Barnich 
and Troessaert 
\cite{Barnich:2011mi} of vector fields and is 
\begin{align}
	[\xi_1,\xi_2]^a_M
	&=
		[\xi_1,\xi_2]^a
		- \delta_{\xi_1} \xi_2^a
		+ \delta_{\xi_2} \xi_1^a
	,
\end{align}
where $\delta_{\xi_1}\xi_2^a$ denotes the change in the vector component $\xi_2^a$ induced by the diffeomorphism $\xi_1$.
Supertranslations and superrotations acting on the metric then form a closed algebra under 
the modified bracket.

For example, given a pair of vector fields $\xi_i$ ($i=1,2$) that generate a supertranslation $f_i$ and
a superrotation $Y_i$, one can show that
\begin{align}
	[\xi_1,\xi_2]_M = \xi_3,
\end{align}
where $\xi_3$ is a vector field that generates both a supertranslation $\hat f$ and a 
superrotation $\hat Y$ given by
\begin{align}
	\hat f
	&=
		\frac{1}{2} f_1 D_AY^A_2
		- \frac{1}{2} f_2 D_AY^A_1
		+ Y_1^A D_A f_2
		- Y_2^A D_A f_1
	,\\
	\hat Y^A &=
		Y_1^B D_B Y_2^A
		- Y_2^B D_B Y_1^A
	.
\end{align}
A derivation of the above result is given in appendix \ref{app:modified}.
We note that this is the same as for the BMS$_4$ algebra at null infinity \cite{Barnich:2011mi}.

Another important ingredient that plays a central role in this work is dual supertranslation,
	which is a new set of asymptotic symmetries of gravity that has recently been uncovered \cite{Godazgar:2018qpq}.
Interestingly, dual supertranslations are not diffeomorphisms of any kind \cite{Kol:2019nkc}, and they have a natural interpretation as the magnetic dual
	of the standard BMS supertranslation
	\cite{Godazgar:2018dvh,Godazgar:2018qpq,Godazgar:2019dkh,Kol:2019nkc}.
In electromagnetism, magnetic large gauge symmetry is tied to the complexification of the large gauge transformation charge (see \cite{Strominger:2015bla} for instance).
Similarly in gravity, the appearance of dual supertranslation can be understood as the complexification of the BMS charge.
Just like the BMS supertranslation charge $Q_f^{\mI^+}$ can be written as the real part of a complex Weyl scalar,
\begin{align}
	Q_f^{\mI^+}
	&=
		\frac{1}{4\pi}\int_{\mI^+_-} d^2z\sqrt\gamma f(z,\zb)\Re\left[\Psi^0_2(u,z,\zb)\right]
	,
\end{align}
the dual supertranslation charge $\tilde Q_f^{\mI^+}$ is associated to its imaginary part \cite{Godazgar:2018qpq,Kol:2019nkc},
\begin{align}
	\tilde Q_f^{\mI^+}
	&=
		\frac{1}{4\pi}\int_{\mI^+_-} d^2z\sqrt\gamma f(z,\zb)\Im\left[\Psi^0_2(u,z,\zb)\right]
	.
\end{align}
A prime example of a spacetime with a non-trivial global dual supertranslation charge is the Taub-NUT spacetime \cite{Taub:1950ez,Newman:1963yy},
	which has been studied in detail in the context of dual supertranslation in \cite{Kol:2019nkc}.
There are also examples of asymptotically flat spacetimes with bulk dust configurations that lead to a non-trivial dual supertranslation at the null infinity,
	see section III.D of \cite{Satishchandran:2019pyc}.

More recently, it has been demonstrated by \cite{Godazgar:2020gqd,Godazgar:2020kqd} that dual supertranslation charges (or dual diffeomorphism charges in general)
	can be computed using covariant phase space formalism in first-order formalism of gravity with the Holst action \cite{Holst:1995pc}.
In the next section, we employ this method to compute the dual supertranslation charge on the future Schwarzschild horizon.
This dual charge is then used along with the standard horizon supertranslation charge to compute the Dirac bracket algebra of horizon charges.


\section{Horizon charges} \label{sec:charge}
We will now  construct the supertranslation charges on the future horizon $\mH^+$ assuming 
smoothness of the supertranslation parameter $f$. 

Following \cite{Hawking:2016sgy}, let us define $\Sigma$ to be a spacelike hypersurface extending from 
a section of $\mI^+$ to a section of the horizon $\mH^+$.
A charge $Q^\Sigma$ associated with $\Sigma$ breaks into two parts, one being on the horizon 
and the other on null infinity. These two parts of $Q^\Sigma$ correspond to the two components of 
the boundary of $\Sigma$, $\partial\Sigma$.
\begin{align}
	Q^\Sigma
	&=
		Q^\H
		+ Q^{\mI^+}
	\label{Q}
	.
\end{align}
In \cite{Godazgar:2020gqd,Godazgar:2020kqd}, the authors provide a formula for the 
(possibly non-integrable) variation of
	electric and magnetic charges associated with a vector field $\xi$. The metric is varied 
	inducing
	a variation of the connection $1$-form $\omega^{\alpha\beta}$ of $\delta\omega^{\alpha\beta}$
\begin{align}
	\slashed \delta Q^\Sigma_E 
	&=
		\frac{1}{16\pi} \ep_{\abcd}
			\int_{\pS}
			(i_\xi E^\c)\delta\w^{\ab}\wedge E^\d
	,\\
	\slashed \delta Q^\Sigma_M
	&=
		\frac{1}{8\pi}\int_{\pS}
			(i_\xi E^\alpha)\delta\w_\ab \wedge E^\beta,
\end{align}
Each of these break into two contributions $\slashed \delta Q^{\mH^+}$ and $\slashed \delta Q^{\mI^+}$.
On the horizon, there is an advanced time coordinate $v$ and
the horizon contributions at time $v_0$ take the form
\begin{align}
	\slashed \delta Q^\H_E 
	&=
		\frac{1}{16\pi} \ep_{\abcd}
			\int_{\, \mH^+_{v_0}}
			(i_\xi E^\c)\delta\w^{\ab}\wedge E^\d
	,\\
	\slashed \delta Q^\H_M
	&=
		\frac{1}{8\pi}\int_{\, \mH^+_{v_0}}
			(i_\xi E^\alpha)\delta\w_\ab \wedge E^\beta
			.
	\end{align}
Throughout this paper, we will take the viewpoint that the black hole ultimately evaporates. Therefore,
although there is a future boundary $\H$ to the horizon, we assume that there is no contribition 
to the charge there. If an horizon has a future end-point, in classical general relativity
it must be singular. We presume, in conformity with common practice, that this is not an issue and that
quantum phenomena will take care of matters. 
We therefore take $\p\H \equiv \mH^+_-$, the past endpoint of the horizon, and ignore all possible 
contributions of $\mH^+_+$, the future endpoint of the horizon.\footnote{
For eternal black holes, one should add boundary degrees of freedom on $\mH^+_+$ such that they cancel the contribution of $\mH^+_+$ to the integral,
	since $\mH^+_+$ is not a genuine part of the boundary $\pS$.
See \cite{Geiller:2017whh,Geiller:2017xad,Speranza:2017gxd,Hosseinzadeh:2018dkh,Freidel:2018fsk} for a discussion of electromagnetism on $\mI^+$.
}

Expressions for the horizon contributions in Bondi coordinates are derived in appendix \ref{app:charges}.
Taking $\xi$ to be the supertranslation vector field
\begin{align}
	\xi = f \p_v - \frac{1}{2} D^2 f\p_r + \frac{1}{r}D^A f\p_A
	\label{STxi}
	,
\end{align}
we obtain the horizon supertranslation charge $\QE$ from \eqref{aQE} and the dual supertranslation charge $\QM$ from \eqref{aQM} to be
\begin{align}
	\QE
	&=
		\frac{M}{8\pi}
			\int_{\mH^+_{v_0}} d^2\x \sqrt\gamma\bigg[
				D^A \left(
					\frac{f}{M}h_{vA}
					+ (D_Af)h_{vr}
				\right)
				\\&\hspace{3cm}
				- (D^A f)\p_rh_{vA}
				+ 2fh_{vv}
				+ (D^2 f) h_{vr}
			\bigg]
		\label{QE}
	,\\
	\QM
	&=
		-\frac{1}{32\pi M}\int_{\mH^+_{v_0}} d^2\x\sqrt\gamma
			(D^B f) \ep_A{}^CD^Ah_{BC}
		\label{QM}
	.
\end{align}
$\ep^{AB}$ is the alternating tensor on the unit 2-sphere and take $\ep^{\t\phi} = \frac{1}{\sin\t}$.

For smooth functions everywhere, we can discard total derivatives in the integrand, and the supertranslation charge is then in exact agreement with that of \cite{Hawking:2016sgy}.
After residual gauge fixing and using a combination of the constraints on $\mH^+$, the supertranslation charge simplifies to the expression
\begin{align}
	\QE
	&=
		\frac{1}{16\pi M}\int_\H dv\,d^2\x \sqrt \gamma f(\x) D^A D^B \sigma_{AB}
	,
\end{align}
where $\sigma_{AB}=\frac{1}{2}\p_v h_{AB}$ is the conjugate momentum of $h_{AB}$.
The integral over the advanced time parameter $v$ is taken from $\mH^+_-$ to $v_0$.
The phase space of the horizon $\H$ has the Dirac bracket \cite{Hawking:2016sgy},
\begin{align}
	\{\sigma_{AB}(v,\Omega),h_{CD}(v',\Omega')\}_D=32\pi M^2\gamma_{ABCD}\delta(v-v')\delta(\Omega-\Omega')
	\label{bracket}
	,
\end{align}
where $\gamma_{ABCD} \equiv \gamma_{AC}\gamma_{BC} + \gamma_{AD}\gamma_{BC} - \gamma_{AB}\gamma_{CD}$
	is proportional to the DeWitt metric \cite{DeWitt:1967yk}.

Since we can integrate by parts freely without having to worry about boundary terms, we can move all covariant derivatives to act on $f$.
As such, we can now identify the integrable horizon supertranslation charge $\iQE$ and dual supertranslation charge $\iQM$ as
\begin{align}
	\iQE
	&\equiv
		\frac{1}{16\pi M}\int_\H dv\,d^2\x \sqrt \gamma\, (D^BD^Af) \sigma_{AB}
	\label{iQE}
	,\\
	\iQM
	&\equiv
		-\frac{1}{32\pi M}\int_{\mH^+_-} d^2\x \sqrt \gamma\, (D^BD^Af) \ep_A{}^Ch_{BC}
	\label{iQM}
	.
\end{align}
Notice that in this form, the dual supertranslation charge is related to supertranslation charge
	by the twisting procedure $h_{AB} \to \ep_A{}^C h_{CB}$ proposed 
	in \cite{Godazgar:2018dvh,Godazgar:2019dkh}.
When we have smooth functions 
	everywhere, $\QE=\iQE$ and $\QM=\iQM$, i.e$.$ the charges are integrable.
	

\section{Supertranslation charge with poles on the complex plane} \label{sec:Qpole}
We now extend the construction of previous section to allow for the possibility that the 
supertranslation parameters, $f$, have simple poles.

The easiest way to explore this possibility is  to use complex stereographic coordinates 
$(z,\zb)$, defined as
\begin{align}
	z = e^{i\phi}\tan\frac{\t}{2}
	,\qquad
	\zb = e^{-i\phi}\tan\frac{\t}{2}
	,
	\label{z}
\end{align}
where $\t$ and $\phi$ are the standard spherical coordinates on a unit sphere.
The metric on the unit sphere in these coordinates is $\gzz = \frac{2}{(1+z\zb)^2}$, $\gamma_{zz}=\gamma_{\zb\zb}=0$.
The integration measure on the sphere is 
\begin{align}
	d^2\x \sqrt\gamma
	&=
		d^2z \sqrt\gamma
	,\ \ \ {\rm with}\ \ \ 
	d^2z\equiv i dz\wedge d\zb
	,\ \ \ {\rm and} \ \ \ 
	\sqrt\gamma = \gzz
	.
\end{align}
The notation has been organized such that $d^2z$ is real.
The alternating tensor is defined such that $\ep_{z\zb}=i\sqrt{\gamma}$.
The only non-vanishing Christoffel symbols are ${}^{(2)}\Gamma^z_{zz} = \frac{-2\zb}{1+z\zb}$ and ${}^{(2)}\Gamma^\zb_{\zb\zb}=\frac{-2z}{1+z\zb}$.

Let us compute the supertranslation charge $\QE$ when $f(z,\zb)$ has a pole at some complex coordinate $w$, that is, $f=\frac{1}{z-w}$.
After fully fixing the residual gauge freedom on $\mH^+$, as in \cite{Hawking:2016sgy},  we have
\begin{align}
\begin{split}
	h_{vv} &= h_{vA} = 0
	,\\
	h_{vr}
	&=
		\frac{1}{4M^2} [D^2-1]^{-1}D^BD^C h_{BC}
	,\\
	\p_r h_{vA}
	&=
		-\frac{1}{4M^2}D_A [D^2-1]^{-1}D^BD^C h_{BC} + \frac{1}{4M^2}D^B h_{AB}
	,
	\label{gaugefixing}
\end{split}
\end{align}
and the supertranslation charge \eqref{QE} takes the form
\begin{align}
	\QE
	&=
		\frac{M}{8\pi}\int_{\p \mH^+} d^2z\sqrt{\gamma}
		\left(
		- (D^A f) \frac{1}{4M^2}D^B h_{AB}
		+ 2D_A(D^A fh_{vr})
		\right)
	.\label{Q7}
\end{align}
In obtaining this we have used \eqref{gaugefixing} for $D_A h_{vr} + \p_r h_{vA}$.
Now consider the total derivative term $D_A(D^A f h_{vr})$.
For $f=\frac{1}{z-w}$ we have,
\begin{align}
	\int d^2z\sqrt{\gamma} \, D^A ((D_A f) h_{vr})
	&=
		i \int dz\wedge d\zb\, (\p_\zb ( h_{vr}\p_z f) + \p_z (h_{vr}\p_\zb f ))
	\\ &=
		-i \oint_w dz \,h_{vr}\p_z f 
		+i \oint_w d\zb\, h_{vr}\p_\zb f 
	\\ &=
		-2\pi \p_zh_{vr}\big|_{z=w}
		\label{dzhvr}
	.
\end{align}
In the second line, the contour is a small circle taken counter-clockwise around $z=w$.
The second term on the r.h.s$.$ of the second line vanishes because $f=\frac{1}{z-w}$ satisfies the identity\footnote{
	Note that we normalize $\delta^2(z-w)$ as a real density, so $1=\int d^2z\,\delta^2(z-w)=\int {\bf{\ep}} \frac{1}{\sqrt\gamma}\delta^2(z-w)$,
	where ${\bf{\ep}}=d^2z\sqrt\gamma$ is the volume form on the unit sphere.
}
\begin{align}
	\p_\zb f = 2\pi\delta^2(z-w)
	.
\end{align}
The contour of $\oint_w d\zb$ is a small circle around $z=w$ and so does not pick up any contribution 
from the delta-function.
In the first term of the second line since $\p_z f=\frac{-1}{(z-w)^2}$,
	there is a contribution proportional to $\p_z h_{vr}$ evaluated at $w$, which is the result  \eqref{dzhvr}.
Substiuting in the expression \eqref{gaugefixing} for $h_{vr}$ we find
\begin{align}
	\QE
	&=
		-\frac{1}{16\pi M}\int_\H dv\,d^2z\sqrt{\gamma}\,
			(D^A f) D^B \sigma_{AB}
		- \frac{1}{4M}\int_{-\infty}^\infty dv D_z [D^2-1]^{-1}D^BD^A \sigma_{AB}\bigg|_{z=w}
	.
	\label{Q0}
\end{align}
Partial integration of the first term gives
\begin{align}
	\QE
	&=
		\frac{1}{16\pi M}\int_\H dv\,d^2z\sqrt{\gamma}\,
			(D^BD^A f) \sigma_{AB}
		- \frac{1}{4M}\int_{-\infty}^\infty dv D_z [D^2-1]^{-1}D^BD^A \sigma_{AB}\bigg|_{z=w}
	.
	\label{Q00}
\end{align}
In \eqref{Q00}, the first term vanishes since 
\begin{align}
	\int d^2z\sqrt{\gamma}\, D^B (\sigma_{AB} D^A f)
	&=
		\int d^2z
		\left(
			\p_\zb (\sigma_{zz} D^z f)
			+ \p_z (\sigma_{\zb\zb} D^\zb f)
		\right)
	\\ &=
		-i\oint_w dz \guzz \sigma_{zz}\p_\zb f 
		+i\oint_w d\zb \guzz \sigma_{\zb\zb}\p_z f 
	\\ &= 0 \label{413}
	.
\end{align}
In obtaining \eqref{413}  we have again used  $\p_\zb f = 2\pi \delta^2(z-w)$,
 the contour of $\oint_wdz$ is a circle around $w$,
	and $\sigma_{\zb\zb}\p_z f=-\frac{1}{2}(z-w)^{-2}\p_vh_{\zb\zb}$ does not have poles in $\zb$.

We recognize the first term in \eqref{Q00} for general $f$ to be the integrable supertranslation 
charge $\iQE$ \eqref{iQE}.
Thus, we find that a pole in $f$ leads $\iQE$ to acquire a non-integrable part $\nQE$,
\begin{align}
	\QE = \iQE + \nQE
	,
\end{align}
where $\iQE$ is given by \eqref{iQE}, and
\begin{align}
	\nQE
	&=
		- \frac{1}{4M}\int_{-\infty}^\infty dv D_z [D^2-1]^{-1}D^BD^A \sigma_{AB}\bigg|_{z=w}
	.
	\label{nQE}
\end{align}
This splitting into integrable and non-integrable parts is, of course, not unique (see for instance \cite{Godazgar:2020kqd}).
Our choice is justified as
	firstly $\iQE$ is the horizon supertranslation charge in the absence of poles in $f$, and
secondly $\nQE$ has zero Dirac bracket with both $\iQE[g]$ and $\iQM[g]$ and so  carries no degrees of freedom.
We encountered the first observation at the end of section \ref{sec:charge} and
we will demonstrate second in appendix \ref{app:nQE}.

\section{Dirac bracket between charges}\label{sec:bracket}

We now compute the Dirac bracket $\{\iQE,\iQM[g]\}_D$, where $f=\frac{1}{z-w}$ and $g$ is 
assumed to be smooth.
This bracket probes central terms of the algebra of charges. To see this note
 that the charges have the expansions,
\begin{align}
	\QEfull
	&=
		Q_f^{(h=0)}
		+ \iQE
		+ \O(h^2)
	,\\
	\QMfull[g]
	&=
		\tilde Q_g^{(h=0)}
		+ \iQM[g]
		+ \O(h^2)
	,
\end{align}
where $Q_f^{(h=0)}$ and $\tilde Q_g^{(h=0)}$ are the constant charges of the background metric and hence do not carry degrees of freedom. This gives,
\begin{align}
	\{\QEfull,\QMfull[g]\}_D
	&=
		\underbrace{\{\iQE,\iQM[g]\}_D}_{\text{constant}}
		+ \O(h)
	.
\end{align}
The constant term corresponds to the central charge of the charge algebra.

Now let us compute $\{\iQE,\iQM[g]\}_D$, with $f=\frac{1}{z-w}$ and $g$ smooth.
Using the expressions \eqref{iQE} and \eqref{iQM} and applying \eqref{bracket}, we obtain
\begin{align}
	\left\{\iQE, \iQM[g]\right\}_D
	&=
		\frac{-1}{2(16\pi M)^2}\left\{
			\int_\H dv\,d^2z\sqrt{\gamma}\, (D^BD^A f) \sigma_{AB}
			,
			\int_\s d^2z\sqrt{\gamma}\, (D^E D^C g) \ep_E{}^D h_{CD}
		\right\}_D
	\nonumber\\ &=
		-\frac{1}{16\pi}
			\int_\s d^2z\sqrt{\gamma}\, (D^BD^A f) (D^E D^C g) \ep_E{}^D
			\gamma_{ABCD} \label{55}
		.
\end{align}
Rearranging $D^B$ in \eqref{55} results in 
\begin{align}
	\left\{\iQE, \iQM[g]\right\}_D
	&=
		-\frac{1}{16\pi}
			\int_\s d^2z\sqrt{\gamma}
			\bigg(
			D^B\left((D^A f) (D^E D^C g) \ep_E{}^D\gamma_{ABCD}\right)
			- (D^A f) (D^B D^E D^C g) \ep_E{}^D\gamma_{ABCD}
			\bigg)
	.
\end{align}
Substituting in the expressions for $\ep_A{}^B$ and $\gamma_{ABCD}$, we can see that the first term is zero,
\begin{align}
	\int_\s d^2z\sqrt{\gamma}\, D^B\left((D^A f) (D^E D^C g) \ep_E{}^D\gamma_{ABCD}\right)
	&=
		\int_\s d^2z\, \p_\zb\left((D^z f)( D^\zb D^\zb g )\ep_\zb{}^\zb\gamma_{zz\zb\zb}\right)
		\nonumber\\&\quad
		+ \int_\s d^2z\, \p_z \left((D^\zb f )(D^z D^z g) \ep_z{}^z\gamma_{\zb\zb zz}\right)
	\nonumber\\ &=
		-2\oint_w dz\, (\p_\zb f) (D^\zb D^\zb g) \gzz
		- 2\oint_w d\zb\,(\p_z f) (D^z D^z g) \gzz
	\nonumber\\ &=
		0
		.
\end{align}
In obtaining this result, we have used the fact that the $\oint_w dz$
integral vanishes since its contour is a circle around $w$ and does 
not intersect the singularity of the delta 
function $\p_\zb f = 2\pi\delta^2(z-w)$,
	and the $\oint_w d\zb$ integral vanishes since $(\p_z f) (D^z D^z g) \gzz$ does not have
	 a pole in $\zb$.
We obtain
\begin{align}
	\left\{\iQE, \iQM[g]\right\}_D
	&=
		\frac{1}{8\pi}
			\int_\s d^2z\sqrt{\gamma}\,\gzz^2
			\left (
			(D^z f) (D^z D^\zb D^\zb g) \ep_\zb{}^\zb
			+ (D^\zb f)( D^\zb D^z D^z g) \ep_z{}^z
			\right )
	\\ &=
		\frac{-i}{8\pi}
			\int_\s d^2z
			\left(
				(\p_\zb f) D^zD_z^2 g
				- (\p_z f) D^\zb D_\zb^2 g
			\right)
		.
	\\ &=
		\frac{-i}{8\pi}
			\int_\s d^2z
			\guzz
			\bigg(
				(\p_\zb f) [D_\zb, D_z]D_z g
				+ (\p_\zb f) D_z D_\zb D_z g
				\nonumber\\&\qquad\qquad\qquad\qquad
				- (\p_z f) [D_z, D_\zb] D_\zb g
				- (\p_z f) D_\zb D_z D_\zb g
			\bigg)
		.
\end{align}
The commutators are $[D_\zb,D_z]D_z g = \gzz D_z g$ and $[D_z,D_\zb]D_\zb g=\gzz D_\zb g$.
Thus we have
\begin{align}
	\left\{\iQE, \iQM[g]\right\}_D
	&=
		\frac{-i}{8\pi}
			\int d^2z
			\left(
				(\p_\zb f) D_z g
				- (\p_z f) D_\zb g
				+ (\p_\zb f) D_z D_\zb D^\zb g
				- (\p_z f) D_\zb D_z D^z g
			\right)
		.
\end{align}
For the last two terms in the parentheses, we have used $\guzz$ to purposely raise the index of the first derivative acting on $g$.
This allows us to write the third covariant derivatives acting on $g$ as partial derivatives,
\begin{align}
\left\{\iQE, \iQM[g]\right\}_D
	&=
		\frac{-i}{8\pi}
			\int d^2z
			\left(
				(\p_\zb f) D_z g
				- (\p_z f) D_\zb g
				+ (\p_\zb f) \p_z D_\zb D^\zb g
				- (\p_z f) \p_\zb D_z D^z g
			\right)
	.
\end{align}
Now we partial integrate all $\p_A f$'s inside the parentheses.
Only the boundary terms survive since partial derivatives commute and
\begin{align}
	D_\zb D^\zb g - D_z D^z g
	&=
	\guzz(\p_\zb \p_z g - \p_z \p_\zb) g
	=
	0
	.
\end{align}
Therefore, we have via Stokes' theorem,
\begin{align}
\left\{\iQE, \iQM[g]\right\}_D
	&=
		\frac{-i}{8\pi}
			\int d^2z
			\left(
				\p_\zb (f D_z g)
				- \p_z (f D_\zb g)
				+ \p_\zb( f \p_z D_\zb D^\zb  g)
				- \p_z(f \p_\zb D_z D^z g)
			\right)
	\\ &=
		-\frac{1}{8\pi}
			\oint_w
			\left(
				dz \frac{(D_z g+\p_zD_\zb D^\zb g)}{z-w}
				+ d\zb \frac{ (D_\zb g+\p_\zb D_z D^z g)}{z-w}
			\right)
		.
\end{align}
The $\oint_w d\zb$ integral vanishes due to the absence of $\zb$-poles.
Now observe that we can use $[D_\zb,D_z]D_z g = \gzz D_z g$ to simplify
\begin{align}
	D_zg + \p_z D_\zb D^\zb g
	&=
		D_zg + D_z D_\zb D^\zb g
	\\ &=
		D_zg + \guzz D_z D_\zb D_z g
	\\ &=
		D_zg + \guzz [D_z, D_\zb] D_z g + \guzz D_\zb D_z D_z g
	\\ &=
		D^z D_z D_z g.
\end{align}
and write
\begin{align}
\left\{\iQE, \iQM[g]\right\}_D
	&=
		-\frac{1}{8\pi}
			\oint_w
				dz \frac{D^z D_z D_z g}{z-w}
\end{align}
The residue theorem then gives
\begin{align}
	\left\{\iQE, \iQM[g]\right\}_D
	&=
		-\frac{i}{4} D^z D_z^2 g\bigg|_{z=w}
		.
		\label{final}
\end{align}

\section{Another approach to the computation of the central term}\label{sec:bracket2}

The result for the central term is new and has important implications. We will now reproduce 
the central term of the previous section using a completely different method.

We start from our expression \eqref{iQM} for the integrable variation $\iQM$ of dual supertranslation charge, which reads
\begin{align}
	\iQM[g]
	&=
		-\frac{1}{32\pi M}\int_{\mH^+_-} d^2z \sqrt \gamma\, (D^BD^Ag) \ep_A{}^Ch_{BC}
		\label{iQMnotes}
	,
\end{align}
and invoke equation (3.4) in the work of Barnich and Troessaert \cite{Barnich:2011mi},
\begin{align}
	\{\QEfull,\QMfull[g]\}_D
	&=
		\delta_f \QMfull[g]
	,
\end{align}
where $\delta_f \tilde Q_g$ denotes taking the expression \eqref{iQMnotes} for 
$\iQM[g]$ and replacing $h_{AB}$ with a diffeomorphism constructed from $f$ with $f$ being only 
dependent on $z$ and $\bar z$.
A general diffeomorphism is of the form $h_{ab}=\nabla_a\xi_b + \nabla_b\xi_a$. Let $\xi_a=\partial_af$.
Then restricting $h_{ab}$ to the sphere gives
\begin{align}
	h_{BC} \ \to \  2M(2D_BD_C f - \gamma_{BC} D^2 f)
	.
\end{align}
This leads to the expression
\begin{align}
	\{\QEfull,\QMfull[g]\}_D
	&=
		-\frac{1}{16\pi}\int d^2z \sqrt \gamma\,
			(D^BD^Ag) \ep_A{}^C(2D_BD_C f - \gamma_{BC} D^2f)
	\\ &=
		-\frac{1}{8\pi}\int d^2z \sqrt \gamma\,
			(D^BD^Ag) \ep_A{}^CD_BD_C f
		+ \frac{1}{16\pi}\int d^2z \sqrt \gamma\,
			(D^BD^Ag) \ep_{AB} D^2f
		.
\end{align}
The second term on the r.h.s$.$ is zero, since $D^BD^Ag$ is symmetric and $\ep_{AB}$ is antisymmetric.
We are just left with the first term,
\begin{align}
	\{\QEfull,\QMfull[g]\}_D
	&=
		-\frac{1}{8\pi}\int d^2z \sqrt \gamma\,
			(D^BD^Ag) \ep_A{}^CD_BD_C f.
\end{align}
Rewrite this as the sum of two terms
\begin{align}
	\{\QEfull,\QMfull[g]\}_D
	&=
		-\frac{1}{8\pi}(X+Y)
	\label{12}
	,
	\end{align}
	with
	\begin{align}
	X
	&\equiv
		\int d^2z \sqrt \gamma
			D_B\left((D^BD^Ag) \ep_A{}^CD_C f\right)
	,\\
	Y
	&\equiv
		- \int d^2z \sqrt \gamma
			(D^2D^Ag) \ep_A{}^CD_C f
	.
\end{align}
$X$ is of the form of an integral over the sphere of the divergence of a vector field $V^A$ on the sphere.
So,

\begin{align}
	\int d^2z\sqrt\gamma D_B V^B
	&=
		i\int dz\wedge d\zb \,\gzz (D_z V^z + D_\zb V^\zb)
	\\ &=
		i\int dz\wedge d\zb \,(\p_z V_\zb + \p_\zb V_z)
	\\ &=
		i\oint_w d\zb\, V_\zb
		- i\oint_w dz\, V_z
	\label{DV}
	.
\end{align}
In the second line, 
we have used the fact that
	the only non-vanishing Christoffel symbols are $\Gamma^z_{zz}$ and $\Gamma^\zb_{\zb\zb}$ to write $D_z V_\zb=\p_zV_\zb$ and $D_\zb V_z=\p_\zb V_z$.
Finally we  use Stokes' theorem to write $X$ as
\begin{align}
	X
	&=
		i\oint_w d\zb (D_\zb D^Ag) \ep_A{}^CD_C f
		- i\oint dz (D_zD^Ag) \ep_A{}^CD_C f
	.
\end{align}
Everything is smooth except for $f=\frac{1}{z-w}$, so the first term with $\oint d\zb$ never sees a pole in $\zb$ and therefore vanishes.
Writing out the second term while noting that the only non-vanishing components of $\ep_A{}^B$ are $\ep_z{}^z=-\ep_\zb{}^\zb=i$, we obtain
\begin{align}
	X
	&=
		\oint dz (D_zD^zg) \p_z f
		- \oint dz (D_zD^\zb g) \p_\zb f
	.
\end{align}
The second term vanishes since it has $\p_\zb f = 2\pi\delta^2(z-w)$ and the contour never meets $w$.
We can partial integrate the first term and using the residue theorem and $f=\frac{1}{z-w}$ obtain
\begin{align}
	X
	&=
		-\oint dz (\p_z D_zD^zg) f
	\\ &=
		-\oint dz \frac{\p_z D_zD^zg}{z-w}
	\\ &=
		-2\pi i \p_z D_zD^zg|_{z=w}
	\label{app1}
	.
\end{align}
Now we turn to $Y$ in \eqref{12}, which reads
\begin{align}
	Y
	&=
		- \int d^2z \sqrt \gamma
			(D^2D^Ag) \ep_A{}^CD_C f
	\\ &=
		- \int d^2z \sqrt \gamma
			D_C\left((D^2D^Ag) \ep_A{}^C f\right)
		+ \int d^2z \sqrt \gamma
			(D_CD^2D^Ag) \ep_A{}^C f
	.
\end{align}
One quickly see that the second term vanishes as
\begin{align}
	\ep_A{}^CD_CD^2D^Ag
	&=
		\ep^{AC}D_CD^2D_Ag
	\\ &=
		\ep^{AC} D_C[D^2,D_A]g
		+ \ep^{AC} D_C D_A D^2 g
	\\ &=
		\ep^{AC} D_CD_Ag
		+ \ep^{AC} D_C D_A D^2 g
	\\ &=
		0
	,
\end{align}
since both $D_CD_Ag$ and $D_CD_A D^2 g$ are symmetric in $A$ and $C$ and
$[D^2,D_A]g = D_A g$.
So we are left with just
\begin{align}
	Y
	&=
		- \int d^2z \sqrt \gamma
			D_C\left((D^2D^Ag) \ep_A{}^C f\right)
	,
\end{align}
which again is of the form \eqref{DV}, so we can writing the explicit form of $f$ as $\frac{1}{z-w}$, and   $\ep_{z\zb}=-\ep_{\zb z}=i\gzz$ we find
\begin{align}
	Y
	&=
		- i\oint_w d\zb (D^2D^zg) \ep_{z\zb}f
		+ i\oint_w dz (D^2D^\zb g) \ep_{\zb z}f
	\\ &=
		\oint_w d\zb \frac{(D^2D_\zb g)}{z-w}
		+ \oint_w dz \frac{(D^2D_z g) }{z-w}
	.
\end{align}
Explicitly writing $f$ as $\frac{1}{z-w}$, and  $\ep_{z\zb}=-\ep_{\zb z}=i\gzz$.
The first term is zero since there are no poles in $\zb$, and the second term yields the residue at $z=w$,
\begin{align}
	Y
	&=
		2\pi D^2D_z g |_{z=w}
	\label{app2}
	.
\end{align}
Collecting the results \eqref{app1} and \eqref{app2} and plugging them into \eqref{12}, we obtain
\begin{align}
	\{\QEfull,\QMfull[g]\}_D
	&=
		-\frac{1}{8\pi}\left(X+Y\right)
	\\ &=
		-\frac{i}{4}\left.\left(
			- \p_z D_zD^zg
			+ D^2D_z g
		\right)\right|_{z=w}
	.
\end{align}
Simplifying
\begin{align}
	- \p_z D_zD^zg
	+ D^2D_z g
	&=
		- \p_z D^z D_z g
		+ D^2D_z g
	\\ &=
		- D_z D^z D_z g
		+ D_z D^z D_z g
		+ D_\zb D^\zb D_z g
	\\ &=
		D_\zb D^\zb D_z g
	\\ &=
		D^z D_z D_z g
	,
\end{align}
where in the second line we have used $\p_z D^z D_z g = D_z D_\zb D^\zb g = D_z D^z D_z g$.
This finally leads to
\begin{align}
	\{\QEfull,\QMfull[g]\}_D
	&=
		-\frac{i}{4}D^z D_z^2 g|_{z=w}
	.
	\label{final2}
\end{align}
This is in complete agreement with our earlier result \eqref{final} for the infinitesimal bracket $\{\iQE,\iQM[g]\}_D$.

What are the implications of this central term? It is usually understood that this is indicative of an anomaly in the theory which must be cancelled in order for the theory to make sense.
In order to understand how to remove the central term in the supertranslation algebra we will 
first take a look at the simpler case of the electromagnetic charges of large gauge transformation 
which is discussed in the next section.


\section{Electromagnetism}\label{sec:em}

 Consider now electromagnetic soft charges on the Schwarzschild horizon. Our discussion  
is parallel to the case of future null infinity $\mI^+$ since both $\H$ and $\mI^+$ are null 
hypersurfaces. We refer the reader to \cite{Hosseinzadeh:2018dkh,Freidel:2018fsk} for a treatment of the 
electromagnetic case on $\mI^+$.

Just like the BMS charges, the electromagnetic charges split into the $\H$ and $\mI^+$ 
contributions \eqref{Q}.
Horizon contributions to the (soft) electric and magnetic charges are given by
\begin{align}
	\eQE
	&=
		\int_{\H} d\lambda\wedge *F
	,\\
	\eQM
	&=
		\int_{\H} d\lambda\wedge F
	,
\end{align}
where $\l$ is an arbitrary function on the sphere.
We use the curly letter $\mQ$ to distinguish these charges from the diffeomorphism charges.

We can write these charges as integrals over the null surface $\H$ subject to the same boundary 
conditions as described in section three. 
In the complex coordinates \eqref{z}
\begin{align}
	\eQE
	&=
		-i\int_{\H} dv\,d^2z \left (\p_\zb \l(*F)_{vz}-\p_z\l(*F)_{v\zb}\right )
	\\ &=
		-\int_{\H} dv\,d^2z \left (F_{vz}\p_\zb \l+F_{v\zb}\p_z\l\right )
	,\\
	\eQM
	&=
		-i\int_{\H} dv\,d^2z \left (F_{vz}\p_\zb \l-F_{v\zb}\p_z\l\right )
	.
\end{align}
Alternatively we can write the charges as integrals over section of the horizon at some instant 
of advanced time $v$. 
In the temporal gauge $A_v=0$, we have $F_{vz} = \p_v A_z$ and
\begin{align}
	\eQM
	&=
		i\int_{\s} d^2z \left (A_z\p_\zb \l-A_\zb\p_z\l\right )
	.
\end{align}
The relevant Dirac bracket is \cite{Freidel:2018fsk} (see \cite{He:2014cra,Strominger:2017zoo} for details on the symplectic structure),
\begin{align}
	\{\eQE,A_z\}_D = -\p_z \lambda
\end{align}
using which we obtain
\begin{align}
	\{\eQE,\eQM[\sigma]\}_D
	&=
		\int_\s d^2z\sqrt{\gamma}\,
			\ep^{AB}\p_A\lambda \p_B\sigma
	\\ &=
		\int_{S^2} d\lambda\wedge d\sigma
	.
	\label{c8c1}
\end{align}
For $\l$ with singularities in $z$, this gives rise to a central term in the algebra, just as in the case of gravity.

To get rid of the central term in the algebra, one may imagine that there exists a boundary theory on $\mH^+$ whose purpose is to cancel the anomalous contribution suggested by the central charge discussed above.
For this purpose, let us consider a $U(1)\times U(1)$ Chern-Simons theory with two independent 1-form fields $a$ and $\tilde a$ on a null surface $\Sigma$,
\begin{align}
	S = \frac{k}{4\pi} \int_\Sigma a\wedge d\tilde a.
\end{align}
Under an electric large gauge transformation $a$ and $\tilde a$ transform as
\begin{align}
	a&\ \to\ a + d\phi
	,\\
	\tilde a&\ \to\ \tilde a
	,
\end{align}
and under a magnetic large gauge transformation they transform as
\begin{align}
	a&\ \to\ a
	,\\
	\tilde a&\ \to\ \tilde a + d\chi
	.
\end{align}
From the action we find the equations of motion to be $da=0$ and $d\tilde a=0$.
Variation of the action yields
\begin{align}
	\delta S
	&=
		\frac{k}{4\pi}\int_{\Sigma} (\delta a\wedge d\tilde a + a \wedge d\delta \tilde a)
	\\ &=
		\frac{k}{4\pi}\int_{\Sigma} (\delta a\wedge d\tilde a - da \wedge \delta \tilde a)
		+ \frac{k}{4\pi}\int_{\p\Sigma} a\wedge \delta \tilde a
	,
\end{align}
from which we obtain the symplectic potential as,
\begin{align}
	\t(a,\tilde a,\delta a,\delta \tilde a)
	&=
		\frac{k}{4\pi} a\wedge \delta \tilde a
	.
\end{align}
Accordingly, the symplectic current density is
\begin{align}
	\w(a,\tilde a,\delta_1 a,\delta_1\tilde a,\delta_2 a,\delta_2 \tilde a)
	&=
		\frac{k}{4\pi}\left(
			\delta_1a\wedge \delta_2 \tilde a
			- \delta_2a\wedge \delta_1 \tilde a
		\right)
	.
\end{align}
Since there are two types of large gauge transformations, we have two integrable charge variations.
One is the electric charge,
\begin{align}
	\delta \mQ_\phi
	&=
		\int_{\p\Sigma}
			\w(a,\tilde a,\delta a,\delta\tilde a,d\phi,0)
	\\ &=
		-\frac{k}{4\pi}
			\int_{\p\Sigma}
			d\phi\wedge \delta \tilde a
	,
\end{align}
the other is the magnetic charge,
\begin{align}
	\delta \tilde \mQ_{\chi}
	&=
		\int_{\p\Sigma}
			\w(a,\tilde a,\delta a,\delta\tilde a,0,d\chi)
	\\ &=
		\frac{k}{4\pi}
			\int_{\p\Sigma}
			\delta a\wedge d\chi
	.
\end{align}
We can compute the algebra using either one of the variations,
\begin{align}
	\{ \mQ_{\phi},\tilde\mQ_{\chi}\}_D
	&=
		\delta_{\phi} \tilde \mQ_{\chi}
	=
		-\delta_{\chi} \mQ_{\phi}
	,
\end{align}
and one can see that we get the same answer for both cases,
\begin{align}
	\{ \mQ_{\phi},\tilde\mQ_{\chi}\}_D
	&=
		-\frac{k}{4\pi}\int_{\p\Sigma} d\phi\wedge d\chi
		\label{emanom}
	.
\end{align}
The electric-electric and magnetic-magnetic brackets vanish regardless of the presence of poles,
\begin{align}
	\{ \mQ_{\phi},\mQ_{\phi^\prime}\}_D
	&=
		0
		\label{emanom2}
	,\\
	\{ \tilde\mQ_{\chi},\tilde\mQ_{\chi^\prime}\}_D
	&=
		0
		\label{emanom3}
	.
\end{align}
Therefore, one finds the algebra to be exactly parallel to that of standard and dual large gauge
transformation charges on the horizon.
The algebra \eqref{emanom}, \eqref{emanom2} and \eqref{emanom3} tells us that putting a 
$U(1)\times U(1)$ Chern-Simons theory with the proper choice of the level $k$ on the horizon, 
we can get rid of the central term obtained earlier in the standard and dual large gauge transformation
algebra.

Chern-Simons theory is a topological theory and as such is independent of the metric. This 
is how it is possible to have a holographic theory defined on the null surface forming the 
horizon. There is no obstacle to theory being defined on a surface with a degenerate metric. 
A further benefit is that, being independnt of the metric, the theory has vanishing energy-momentum tensor
and so does not affect the spacetime geometry.  

\section{Gravitational Chern-Simons theory}\label{sec:cs}

Gravity in three dimensions is a topological theory. Suppose one starts from the Einstein action, with
or without a cosmological term, and count up the number of physical degrees of freedom at 
each point in spacetime. In $d$-dimensions the metric has $\tfrac{1}{2}d(d+1)$ components. 
Diffeomorphisms, being related to first class constraints generated by a vector fields that subtract out
$2d$ degrees of freedom. The total number of physical degrees of freedom is therefore $\tfrac{1}{2}d(d-3)$.
So in $d=3$ there are no local degrees of freedom. We should therefore expect to find a topological
gravitational theory that is independent of any metric. The Einstein action is not such a construct.
However, Witten \cite{Witten:1989ip} found a Chern-Simons theory that is equivalent to the Einstein theory provided some 
of its fields are identified with the metric. The Chern-Simons theory is independent of any metric
and can therefore be formulated consistently on null surfaces where the spacetime metric is degenerate.
In more conventional theories the necessity of using an inverse metric prevents their formulation on null
surfaces.

\subsection{Chern-Simons Actions}

We now briefly describe Witten's gravitational Chern-Simons theory.
The ingredients are a basis of one-forms $e^a=e^a_idx^i$, a connection one-form 
$\omega^{ab}=\omega_i{}^{ab}dx^i$ and a dimensionful real parameter $\lambda$ that is in 
many way analogous to
a cosmological constant. The indices $i,j \ldots$ are spacetime indices 
whereas $a,b \ldots$ are tangent space indices. Spacetime indices never need to be raised or lowered,
however we do need as an extra piece of spacetime structure, the alternating symbol $\epsilon^{ijk}$.
By contrast, tangent space indices are raised or lowered using the Lorentz metric $\eta_{ab}$.
To construct a three-dimensional spacetime, we construct its metric $g_{ij}$ using 
$e_i^ae_j^b\eta_{ab}$ where $\eta_{ab}={\rm diag}(-++)$.  
In Witten's approach to gravity in three spacetime dimensions, this identification is used to conclude 
the equivalence with the Einstein theory. A Chern-Simons theory needs a gauge group ${\bf G}$,
 and in the case
$\lambda=0$, ${\bf G}$ is chosen to be $ISO(2,1)$. If $\lambda<0$, ${\bf G}$ is $SO(3,1)$ 
and if $\lambda>0$, ${\bf G}$ is
$SO(2,2)$. Note that in last case the gauge group can be factorized
as $SO(2,2) \equiv SL(2,{\mathbb R})\otimes SL(2,{\mathbb R})$. The case of $SO(3,1)$ 
cannot be factorized, but it can be regarded as a complex group $SL(2,{\mathbb C})$.

One can write an all encompassing gauge field $A_i=e_i^aP_a + \omega_i^aJ_a$ where
$\omega_i^a=\tfrac{1}{2}\epsilon^{abc}\omega_i{\,}_{bc}$ and $P_a$ and $J_a$ are the
generators of the gauge group. They have the commutation relations
\begin{equation} 
[J_a,J_b]=\epsilon_{abc}J^c,\ \ \ [J_a,P_b]=\epsilon_{abc}P^c,\ \ \ 
[P_a,P_b]=\lambda\epsilon_{abc}J^c \label{eq:CR}
\end{equation}
For arbitrary $\lambda$,  the Killing form is given by
\begin{equation}
\langle J_a,J_b \rangle =0,\ \ \ \langle J_a,P_b \rangle = \eta_{ab},\ \ \ \langle P_a,P_b \rangle = 0.
\label{eq:KFA} 
\end{equation}
However, when $\lambda\ne 0$,  ${\bf G}$ factorises, a second 
Killing form exists
\begin{equation}
\langle J_a,J_b \rangle =\eta_{ab},\ \ \ \langle J_a,P_b \rangle = 0,\ \ \ \langle P_a,P_b \rangle = 
\lambda \eta_{ab}. \label{eq:KFB}
\end{equation}
If $\lambda=0$, the second Killing form is degenerate and not particularly useful.

From these relations we can construct Chern-Simons theory from the general expression
\begin{equation}
I_{CS} = \frac{k}{4\pi}\int \tr\ \bigl(A\wedge dA + \tfrac{2}{3} A \wedge A \wedge A\bigr).
\end{equation}
If $\lambda\ne 0$, we can construct two different actions using the two different Killing forms.
For any value of $\lambda$, we can construct an \lq\lq electric" theory using the
Killing form of (\ref{eq:KFA}). In terms of the differential forms $e^a$ and $\omega^a$
we get
\begin{equation}
I_{electric}  = \frac{k}{2\pi}\int 2e^a\wedge d\omega_a + \epsilon_{abc}  e^a\wedge \omega^a\wedge \omega^c
+\, \tfrac{1}{3}\lambda \epsilon_{abc} e^a\wedge e^b\wedge e^c. \label{eq:actA}
\end{equation}
or, perhaps more conveniently for some of the following calculations, in terms of components
\begin{equation}
I_{electric} =\frac{k}{2\pi} \int \ d^3x\ \epsilon^{ijk}\ e_i^a\,\bigl(2\partial_j\omega_{ka} + \epsilon_{abc}\omega_j^b
\omega_k^c + \tfrac{1}{3}\lambda\epsilon_{abc}e^b_je^c_k\bigr).
\end{equation}
When $\lambda\ne 0$, we can use the alternative Killing form (\ref{eq:KFB}) to construct a different
action, the \lq\lq magnetic" action
\begin{equation}
I_{magnetic} = \frac{\tilde k}{\pi}\int \omega^a\wedge d\omega_a + \tfrac{1}{3}\epsilon_{abc}\omega^a \wedge \omega^b
\wedge \omega^c + \lambda e^a \wedge de_a + \lambda\epsilon_{abc}\omega^a\wedge e^b \wedge e^c. 
\label{eq:actB}
\end{equation}
This too can be more conveniently for practical calculations be written in terms of components as
\begin{equation}
I_{magnetic} = \frac{\tilde k}{\pi}\int \ d^3x\ \epsilon^{ijk}\ \Bigl(\omega_i^a\,\bigl(\partial_j\omega_{ka} + 
\tfrac{1}{3}\epsilon_{abc}\omega_j^b\omega_k^c\bigr) +
\lambda e_i^a\partial_je_{ka} + \lambda\epsilon_{abc}\omega^a_ie^b_je^k_c \Bigr).
\end{equation}

Both the electric and the magnetic action have the same equations of motion and the same
gauge invariance. The equation of motion from variation $e^a$ in the electric action is
\begin{equation} d\omega^a + \frac{1}{2}\epsilon^{abc}\omega_b\wedge\omega_c +
\tfrac{1}{2}\lambda\epsilon_{abc}e^b\wedge e^c =0. \end{equation} 
It is the analog of the Einstein equation and specifies the curvature of the connection 
$\omega^a$. Variation of $\omega^a$ in the electric action
gives
\begin{equation}
de^a + \epsilon^{abc}\omega_b\wedge e_c = 0 \end{equation}
which shows that the connection is torsion-free.
For the magnetic action, it is the variation of $e^a$ that specifies the curvature of the connection 
and the variation of $\omega^a$ that tells us that it is torsion-free. It is in this sense that these
two actions are dual to each other.

The gauge transformations are of two types. The first is labeled by
a tangent-space vector $\rho^a$. 
The gauge variations of $e^a$ and $\omega^a$ are
\begin{equation} 
\delta e_i^a = -\partial_i\rho^a - \epsilon^{abc}\omega_{ib}\wedge \rho_c
\end{equation}
and
\begin{equation}
\delta\omega_i^a = -\lambda\epsilon^{abc}e_{i\,b}\rho_c.
\end{equation}

The second gauge transformation is generated by a second vector $\tau^a$.
The resulting gauge variations are
\begin{equation}
\delta e_i^a = -\epsilon^{abc}e_{ib}\wedge \tau_c
\end{equation}
and
\begin{equation}
\delta \omega_i^a = -\partial_i\tau^a - \epsilon^{abc}\omega_{ib}\wedge \tau_c.
\end{equation}
After recalling that one has dualised the spin connection, one observes that the 
$\tau$-transformations are just local Lorentz rotations. 

The nature of diffeomorphisms is not quite so striaghtforward. Suppose that one has
a diffeomorphism generated by an
infinitesial vector
field $v^i$. The the variation of the components of the basis of $1$-forms is
\begin{align}
	\delta e_i^a = -v^k(\partial_ke_i^a-\partial_ie_k^a) - \partial_i(v^ke_k^a)
\end{align}
Similarly, the variation of the spin connection is
\begin{align}
	\delta \omega_i^a = -v^k(\partial_k\omega_i^a-\partial_i\omega_k^a) 
	- \partial_i(v^k\omega_k^a).
\end{align}
We now see how to find a diffeomorphism in terms of $\rho^a$ and $\tau^a$. 
Setting
\begin{align} \rho^a = v^ke_k^a \ \ \ {\rm and} \ \ \ \tau^a = v^k\omega_k^a \end{align}
reproduces what is expected for the transformations of both $e^a$ and $\omega^a$  
under a diffeomorphism.

\subsection{The Charges}

We now need to find the soft charges resulting from this pair of actions. The calculation
is routine in the covariant phase space formalism. Firstly one performs a variation of the action in
terms of the variation of the fields $\delta e^a$ and $\delta\omega^a$.
The bulk term then gives the usual equations of motion which we have already described. However, there
is also a boundary term, the symplectic potential $\theta$. For our actions we find for the electric case
\begin{align}
	\theta_{electric} = -\frac{k}{\pi}\omega^a\wedge \delta\omega_a
\end{align}
and for the magnetic case
\begin{align}
	\theta_{magnetic} = -\frac{\tilde k}{\pi}\left(\omega^a\wedge\delta\omega_a +\lambda e^a\wedge\delta e_a\right).
\end{align}

Given a symplectic potential, one finds the symplectic form $\Omega$ by carrying out a 
second variation in $\theta$ of the 
fields, $\delta^\prime e^a$ and $\delta^\prime\omega^a$, antisymmetrising over the two variations
and integrating the resultant $2$-form  over a spacelike surface $\Sigma$.   For the electric
action we find
\begin{align}
	\Omega_{electric} =
	-\frac{k}{\pi} \int_\Sigma \delta e^a \wedge \delta^\prime\omega_a + 
	\delta\omega^a\wedge\delta^\prime e_a
\end{align}
and for the magnetic case
\begin{align}
	\Omega_{magnetic} =
	-\frac{2\tilde k}{\pi} \int_\Sigma \delta^\prime\omega^a\wedge\delta\omega_a + 
	\lambda \delta^\prime e^a \wedge \delta e_a.
\end{align}

The charges are now found by setting the second variation  $\delta^\prime e^a$ and $\delta^\prime\omega^a$
to be pure gauge transformations determined by $\rho^\prime$ and $\tau^\prime$. Now substituting
these variations into the symplectic form and using the equations of motion, one finds that the 
integral for $\Omega$ collaspe into boundary terms giving the variation of the charges conjugate to
$\rho^\prime$ and $\tau^\prime$ on $\partial\Sigma$ under the variation of the fields
$\delta e^a$ and $\delta\omega^a$.

For the electric case, we find
\begin{align}
	\delta Q^E_{\rho,\tau} =
	-\frac{k}{\pi} \int_{\partial\Sigma} \tau_a \delta e^a + \rho_a
	\delta\omega^a
\end{align}
and for the magnetic case
\begin{align}
	\delta Q^M_{\rho,\tau} =
	-\frac{2\tilde k}{\pi} \int_{\partial\Sigma} \tau_a \delta\omega^a +
	\lambda\rho_a\delta e^a
	.
	\label{mdQ}
\end{align}
Both of these charges are integrable, and so we will define the charges to be
\begin{align}
	Q^E_{\rho,\tau} = -\frac{k}{\pi}\int_{\partial\Sigma} \ \tau_ae^a+\rho_a\omega^a 
	\label{eQ}
\end{align}
for the electric case and 
\begin{align}
	Q^M_{\rho,\tau} = -\frac{2\tilde k}{\pi}\int_{\partial\Sigma} \tau_a\omega^a + \lambda\rho_ae^a
	\label{mQ}
\end{align}
for the magnetic case. 

A knowledge of the symplectic form allows one to compute the Dirac bracket of various quantities
of importance in the theory. For reasons that will be explained  later, we do this now for just the
electric theory. On the sphere coordinatised by the complex coordinate $z$, the (electric) symplectic form
becomes
\begin{align}
\Omega_{electric} = \frac{ik}{\pi} \int d^2z \bigl(\delta e^a_z\ \delta^\prime\omega_{\bar z\, a}
-\delta e^a_{\bar z}\delta^\prime \omega_{z\, a} - (\delta \leftrightarrow \delta^\prime)\bigr).
\end{align}
From this it follows that only non-trivial Dirac brackets are
\begin{align}
\{e^a_z(z,\bar z),\omega^b_{\bar z}\} = -\frac{i\pi}{k}\eta^{ab} \delta^2(z-z^\prime)
\end{align}
and its complex conjugate.

We can use these expressions to compute the bracket of the charges with the field variables. Modulo 
the equations of motion, these brackets should reproduce the gauge transformations of the fields.
Explicit calculation reveals that
\begin{align}
	&\{Q^E,e_i^a\} = -\partial_i \rho^a - \epsilon^{abc}e_{i\, b}\tau_c - \epsilon^{abc}\omega_{i\, b}\rho_c
	\nonumber,\\
	&\{Q^E,\omega^a_i\} = -\partial_i\tau^a - \epsilon^{abc}\omega_{i\, b}\tau_c 
	-\lambda\epsilon^{abc}e_{i\, b}\rho_c
	,
	\label{eq:QE}
\end{align}
as expected.
Similarly, the brackets of the magnetic charges with $e^a$ and $\omega^a$ are
\begin{align} 
	&\{Q^M,e^a_i\} = 2\frac{\tilde k}{k}\Bigl( -\partial_i \tau^a - \epsilon^{abc}\omega_{i\, b}\tau_c - \lambda\epsilon^{abc}e_{i\, b}\rho_c\Bigr)
	\nonumber,\\
	&\{Q^M,\omega^a_i\} = 2\lambda\frac{\tilde k}{k} \Bigl(-\partial_i\rho^a - \epsilon^{abc}\omega_{i\, b}\rho_c 
	-\epsilon^{abc}e_{i\, b}\tau_c\Bigr)
	.
	\label{eq:QM}
\end{align}
Again, these are gauge transformations but with the role of $\tau$ and $\omega$ interchanged
and rescaled.

\subsection{Charge algebra}

We now compute the charge algebra. From hereon we are going to work exclusively with the electric
theory. One might then wonder what the point of introducing the magnetic theory is. The answer is that 
allows us to find the magnetic charges in a straightforward fashion. Had we not done so, finding the 
magnetic charges would have been an involved, convoluted and obscure process. The magnetic charges
still exist in the electric theory just as electric charges exist in the magnetic theory. However,
one needs to make a choice of symplectic form at some point and we choose the electric picture.   

\subsubsection{Electric-electric bracket}

The bracket between two electric charges is
\begin{align}
	\{Q^E_{\tau,\rho},Q^E_{\tau',\rho'}\}
	&=
		-\frac{k}{\pi}\int_{\p\Sigma} \left(
			\ep^{abc}\left(
				\tau'_b\tau_c
				+ \lambda\rho'_b\rho_c
			\right)e_a
			+ \ep^{abc}\left(
				\tau'_b \rho_c
				- \tau_b \rho'_c
			\right) \w_a
		\right)
		\nonumber\\&\quad
		- \frac{k}{\pi}\int_{\p\Sigma} \left(
			\rho^a d\tau'_a
			+ \tau^a d\rho'_a
		\right)
	.
\end{align}
Recall from \eqref{eQ} that the integrated electric charge takes the form
\begin{align}
	Q^E_{\tau,\rho}
	&=
		-\frac{k}{\pi}\int_{\p\Sigma} \left(
			\tau_a e^a
			+ \rho_a \w^a
		\right)
	.
\end{align}
Comparing this with the result for the bracket, we observe that
\begin{align}
	\{Q^E_{\tau,\rho},Q^E_{\tau',\rho'}\}
	&=
		Q^E_{\tau'',\rho''}
		- \frac{k}{\pi}\int_{\p\Sigma} \left(
			\rho^a d\tau'_a
			+ \tau^a d\rho'_a
		\right)
	,
	\label{bracketEE}
\end{align}
where
\begin{align}
	\tau''^a
	&=
		\ep^{abc}\left(
			\tau'_b\tau_c
			+ \lambda\rho'_b\rho_c
		\right)
		\label{tau''}
	,\\
	\rho''^a
	&=
		\ep^{abc}\left(
			\tau'_b \rho_c
			- \tau_b \rho'_c
		\right)
		\label{rho''}
	.
\end{align}
With $\rho^a=v^i e_i^a$, the central term is zero whenever $\tau^a=0$ or $\tau^a=v^i\w_i^a$.

\subsubsection{Electric-magnetic bracket}\label{sec:EMbracket}

The bracket between electric and magnetic charges can be obtained in two distinct ways since
\begin{align}
	\{Q^E_{\tau,\rho},Q^M_{\tau',\rho'}\}
	&=
		\delta^E_{\tau,\rho} Q^M_{\tau',\rho'}
	=
		- \delta^M_{\tau',\rho'} Q^E_{\tau,\rho}
	,
\end{align}
where here $\delta^E_{\tau,\rho}$ denotes the gauge transformation generated 
by $Q^E_{\rho,\tau}$ given in  (\ref{eq:QE})and $\delta^M_{\tau',\rho'}$ denotes
	the gauge transformation generated by $Q^M_{\rho,\tau}$ given in (\ref{eq:QM}).
These two results must agree.
Let us first compute
\begin{align}
	\delta^E_{\tau,\rho} Q^M_{\tau',\rho'}
	&=
		- \frac{2\tilde k}{\pi}\int_{\p\Sigma}\left(
			\ep^{abc}	\left(
				\tau'_b \tau_c
				+ \lambda \rho'_b \rho_c
			\right) \w_a
			+ \lambda\ep^{abc}\left(
				\tau'_b \rho_c
				- \tau_b \rho'_c
			\right) e_a
		\right)
		\nonumber\\&\quad
		- \frac{2\tilde k}{\pi}\int_{\p\Sigma}\left(
			\tau_a d\tau'^a
			+ \lambda \rho_a d\rho'^a
		\right)
	.
\end{align}
Comparing this to the magnetic charge \eqref{mQ}, we can see that
\begin{align}
	\delta^E_{\tau,\rho} Q^M_{\tau',\rho'}
	&=
		\tilde Q^M_{\tau'',\rho''}
		- \frac{2\tilde k}{\pi}\int_{\p\Sigma}\left(
			\tau_a d\tau'^a
			+ \lambda \rho_a d\rho'^a
		\right)
\end{align}
where $\tau''$ and $\rho''$ are defined in \eqref{tau''} and \eqref{rho''}.

Let us next use \eqref{eq:QM}  to compute $-\delta^M_{\tau',\rho'} Q^E_{\tau,\rho}$.
We obtain
\begin{align}
	- \delta^M_{\tau',\rho'} Q^E_{\tau,\rho}
	&=
		-\frac{2\tilde k}{\pi}\int_{\p\Sigma} \left(
			\ep^{abc}\left(
				\tau'_b\tau_c
				+ \lambda \rho'_b\rho_c
			\right) \w_a
			+ \lambda \ep^{abc} \left(
				\tau'_b\rho_c
				- \tau_b\rho'_c
			\right) e_a
		\right)
		\nonumber\\&\quad
		-\frac{2\tilde k}{\pi}\int_{\p\Sigma} \left(
			\tau_a d\tau'^a
			+ \lambda \rho_a d\rho'^a
		\right)
	.
\end{align}
Observe that this is exactly the same as the expression for $\delta^E_{\tau,\rho} Q^M_{\tau',\rho'}$.
This is a nice consistency check.

Therefore, we conclude that the electric-magnetic charge bracket is
\begin{align}
	\{Q^E_{\tau,\rho},Q^M_{\tau',\rho'}\}
	&=
		\tilde Q^M_{\tau'',\rho''}
		- \frac{2\tilde k}{\pi}\int_{\p\Sigma}\left(
			\tau_a d\tau'^a
			+ \lambda \rho_a d\rho'^a
		\right)
	,
	\label{bracketEM}
\end{align}
with $\tau''$ and $\rho''$ given in \eqref{tau''} and \eqref{rho''}.

\subsubsection{Magnetic-magnetic bracket}\label{sec:MMbracket}

The bracket between two magnetic charges is
\begin{align}
	\{Q^M_{\tau,\rho},Q^M_{\tau',\rho'}\}
	&=
		\delta^M_{\tau,\rho} Q^M_{\tau',\rho'}
	.
\end{align}
Using \eqref{mdQ} and \eqref{eq:QM}, we obtain
\begin{align}
	\{Q^M_{\tau,\rho},Q^M_{\tau',\rho'}\}
	&=
		- \frac{4\lambda\tilde k^2}{\pi k}\int_{\p\Sigma} \left(
			\ep^{abc} \left(
				\tau'_b \tau_c
				+ \lambda \rho'_b \rho_c
			\right) e_a
			+ \ep^{abc} \left(
				\tau'_b \rho_c
				- \tau_b \rho'_c
			\right) \w_a
		\right)
		\nonumber\\&\quad
		- \frac{4\lambda\tilde k^2}{\pi k}\int_{\p\Sigma} \left(
			\rho^a d\tau'_a
			+ \tau^a d\rho'_a
		\right)
	.
\end{align}
Comparing this to the electric charge \eqref{eQ}, we conclude that
\begin{align}
	\{Q^M_{\tau,\rho},Q^M_{\tau',\rho'}\}
	&=
		4\lambda\frac{\tilde k^2}{k^2} Q^E_{\tau'',\rho''}	
		- 4\lambda\frac{\tilde k^2}{\pi k} \int_{\p\Sigma} \left(
			\rho^a d\tau'_a
			+ \tau^a d\rho'_a
		\right)
	.
	\label{bracketMM}
\end{align}
Again, $\tau''$ and $\rho''$ are given in \eqref{tau''} and \eqref{rho''}.
The central term is the same (up to a constant) as that of $\{Q^E,Q^E\}$, so it vanishes 
for supertranslations.

It may be worth noting that there is a relation
\begin{align}
	\{Q^M_{\tau,\rho},Q^M_{\tau',\rho'}\}
	= 4\lambda \frac{\tilde k^2}{k^2} \{Q^E_{\tau,\rho},Q^E_{\tau',\rho'}\}
	.
\end{align}
The factor of $4\lambda$ seems to be just an artifact for a less than optimal choice of scale for 
$Q^M$ (and preceding that for $I_{magnetic}$).
For instance, if we started from $2I_{magnetic}$  we would have $2Q^M$ in place of $Q^M$ and 
this would have  led to having $16\lambda$ in place of the factor $4\lambda$.

\subsection{$e^a$ and $\w^a$ on the horizon}

In this section, we consider putting a gravitational Chern-Simons theory on the future Schwarzschild horizon $\H$,
	and find the solutions of the equations of motion for $e^a$ and $\w^a$.
We observe that the ``cosmological constant'' $\lambda$ is fixed by the equations of motion.

In the context of our work, $g_{ij}$ is the pullback of the four-dimensional metric in advanced Eddington-Finkelstein coordinates to the future Schwarzschild horizon,
\begin{align}
	g_{ij} = 4M^2\begin{pmatrix}
		0&0&0\\
		0&0&\frac{2}{(1+z\zb)^2}\\
		0&\frac{2}{(1+z\zb)^2}&0
	\end{pmatrix},
\end{align}
where $i,j$ span $(v,z,\zb)$.
The ``flat metric'' $\eta_{ab}$ is the Cartan metric $\eta_{ab} = \diag(-1,1,1)$.
They are connected by the ``triad''
\begin{align}
	e_i{}^a
	&=
		2M\begin{pmatrix} 0&0&0\\0&\frac{1}{1+z\zb}&\frac{i}{1+z\zb}\\0&\frac{1}{1+z\zb}&\frac{-i}{1+z\zb}
		\end{pmatrix}
\end{align}
that satisfies
\begin{align}
	e_i{}^ae_j{}^b\eta_{ab} = g_{ij}
	.
\end{align}
We do not have the inverse relation $g^{ij}e_i{}^ae_j{}^b=\eta^{ab}$ because $g_{ij}$ is not invertible.
We can write the above matrix form as collection of one-forms,
\begin{align}
	e^0
	&=
		0
	,\\
	e^1
	&=
		\frac{2M}{1+z\zb}dz+\frac{2M}{1+z\zb}d\zb
		\label{e}
	,\\
	e^2
	&=
		\frac{2iM}{1+z\zb}dz-\frac{2iM}{1+z\zb}d\zb
	,
\end{align}
from which we obtain
\begin{align}
	dz
	&= 
		\frac{1}{4M}(1+z\zb)\left(
			e^1-ie^2
		\right)
	,\\
	d\zb
	&=
		\frac{1}{4M}(1+z\zb)\left(
			e^1+ie^2
		\right)
	,\\
	dz\wedge d\zb
	&=
		\frac{i}{8M^2}(1+z\zb)^2
			e^1\wedge e^2
	.
\end{align}
The spin connection can be obtained using the equations of motion and the anholonomy coefficients
\begin{align}
	de^a
	&=
		-\w^a{}_b\wedge e^b
	=
		\frac{1}{2}c^a{}_{bc}e^b\wedge e^c
	,\\
	\w_{ab}
	&=
		\frac{1}{2}(c_{abc}-c_{bac}-c_{cab})e^c
	,
\end{align}
where we keep in mind that $\w^{bc} = -\w^a\ep_a{}^{bc}$.
The exterior derivative of $e^a$ yields 
\begin{align}
        de^0&=0\\
	de^1
	&=
		2M\frac{(z-\zb)}{(1+z\zb)^2}
		dz\wedge d\zb
	=
		\frac{i}{4M}(z-\zb)
		e^1\wedge e^2
	,\\
	de^2
	&=
		2iM\frac{(z+\zb)}{(1+z\zb)^2}
		dz\wedge d\zb
	=
		-\frac{1}{4M}(z+\zb)
		e^1\wedge e^2
	,
\end{align}
from which we read off
\begin{align}
	c^1{}_{12}=c_{112}=\frac{i}{4M}(z-\zb)
	,\qquad
	c^2{}_{12}=c_{212}=-\frac{1}{4M}(z+\zb)
	,
\end{align}
with all other coefficients vanishing.
Accordingly, the only non-vanishing component of the spin connection is
\begin{align}
	\w_{12}
	&=
		- \frac{i}{4M}(z-\zb)e^1
		+ \frac{1}{4M}(z+\zb)e^2
	\\ &=
		\frac{i\zb}{1+z\zb} dz
		-\frac{i z}{1+z\zb}d\zb
		\label{w}
	.
\end{align}
The only non-vanishing component of the dual $\w^a=\frac{1}{2}\ep^{abc}\w_{bc}$ is thus
\begin{align}
	\w^0
	&=
		-\w_{12}
	=
		-\frac{i\zb}{1+z\zb} dz
		+\frac{i z}{1+z\zb}d\zb
\end{align}
since $\ep^{012}=-1$.

Let us see if this satisfies the other set of equations of motion
\begin{align}
	d\w^a + \frac{1}{2}\ep^{abc}\w_b\wedge \w_c + \frac{\lambda}{2}\ep^{abc}e_b\wedge e_c
	&= 0
	.
\end{align}
Since only $\w^0$ is non-zero, we have $\ep^{abc}\w_b\wedge \w_c=0$.
The only non-vanishing component of $d\w^a$ is
\begin{align}
	d\w^0
	&=
		\frac{2i}{(1+z\zb)^2}dz\wedge d\zb
	,
\end{align}
and the only non-vanishing term of $\frac{\lambda}{2}\ep^{abc}e_b\wedge e_c$ is
\begin{align}
	\frac{\lambda}{2}\ep^{0bc}e_b\wedge e_c
	&=
		-\lambda e^1\wedge e^2
	=
		\frac{8iM^2\lambda}{(1+z\zb)^2}dz\wedge d\zb
	.
\end{align}
Therefore, the above equations of motion boils down to fixing $\lambda$,
\begin{align}
	\lambda = -\frac{1}{4M^2}.	
\end{align}


\subsubsection{Compensating $\tau$-transformation for central term}

We have seen that the electric and magnetic charges satisfy the algebra \eqref{bracketEE}, \eqref{bracketEM} and \eqref{bracketMM}, which reads
\begin{align}
	\{Q^E_{\tau,\rho},Q^E_{\tau',\rho'}\}
	&=
		Q^E_{\tau'',\rho''}
		- \frac{k}{\pi}\int_{\p\Sigma} \left(
			\rho^a d\tau'_a
			+ \tau^a d\rho'_a
		\right)
	,\\
	\{Q^E_{\tau,\rho},Q^M_{\tau',\rho'}\}
	&=
	Q^M_{\tau'',\rho''}
		- \frac{2\tilde k}{\pi}\int_{\p\Sigma}\left(
			\tau_a d\tau'^a
			+ \lambda \rho_a d\rho'^a
		\right)
	,\\
	\{Q^M_{\tau,\rho},Q^M_{\tau',\rho'}\}
	&=
		4\lambda \frac{\tilde k^2}{k^2} Q^E_{\tau'',\rho''}	
		- 4\lambda\frac{\tilde k^2}{\pi k}\int_{\p\Sigma} \left(
			\rho^a d\tau'_a
			+ \tau^a d\rho'_a
		\right)
	,
\end{align}
with the composition $\tau''$ and $\rho''$ given by \eqref{tau''} and \eqref{rho''}.
We want the central terms of this algebra to cancel the central term of supertranslation 
algebra on the Schwarzschild horizon.
Recall that the $\rho$ transformation is related to a diffeomorphism $v^i$ by
\begin{align}
	v^i
	&=
		\left(f, \frac{1}{2M} D^z f, \frac{1}{2M}D^\zb f\right)
	,\\
	\rho^a
	&=
		v^i e_i^a
	\\ &=
		\frac{1}{1+z\zb}\left(
			0
			,\ 
			D^z f + D^\zb f
			,\ 
			i(D^z f-D^\zb f)
		\right)
	.
\end{align}
We demand that, in this Chern-Simons theory, supertranslation is accompanied a compensating Lorentz transformation ($\tau$-transformation)
	given by
\begin{align}
	\tau^a
	&=
		\left(
			\frac{1}{8\tilde k^{1/2}}(D^2+2)f,
			i\sqrt\lambda\rho^2,
			-i\sqrt\lambda\rho^1
		\right)
	.
\end{align}
This leads to the algebra
\begin{align}
	\{Q^E_{\tau,\rho},Q^E_{\tau',\rho'}\}
	&=
		Q^E_{\tau'',\rho''}
	,\\
	\{Q^E_{\tau,\rho},Q^M_{\tau',\rho'}\}
	&=
		 Q^M_{\tau'',\rho''}
		+ \frac{i}{4} (D^z D_z^2 f')|_{z=w}
	,\\
	\{Q^M_{\tau,\rho},Q^M_{\tau',\rho'}\}
	&=
		4\lambda\frac{\tilde k^2}{k^2} Q^E_{\tau'',\rho''}
	.
\end{align}
Observe that the standard supertranslations commute by themselves, the dual supertranslations commute by themselves, but the standard and dual charges have the correct form of central term.
Thus, we see that exactly the form of the central term obtained in sections \ref{sec:bracket} and \ref{sec:bracket2} is reproduced. 
The Chern-Simons theory can then be used to cancel the  anomalous behavior of the 
supertranslation charge algebra in the case that the supertranslation parameter $f$ has a pole.

Finally, we note that the constant $\tilde k$ can also be fixed in terms of $k$ by demanding that the complexified charge algebra is closed up to the central terms.
One may readily check that the complexified charge $\textbf{Q}_{\tau,\rho} = Q^E_{\tau,\rho} + i Q^M_{\tau,\rho}$ satisfies the bracket
\begin{align}
	\{
		\textbf{Q}_{\tau,\rho},
		\textbf{Q}_{\tau',\rho'}
	\}
	&=
		\left[
			1 - 4\lambda\frac{\tilde k^2}{k^2}
		\right]Q_{\tau'',\rho''}^E
		+ 2i Q_{\tau'',\rho''}^M
		- \left.\frac{1}{2}(D^zD_z^2 f')\right|_{z=w}
	.
\end{align}
For this to close up to the central term,
	we demand that the coefficient of $Q^E_{\tau'',\rho''}$ be $2$, which fixes $\tilde k^2 = -\frac{k^2}{4\lambda} = k^2M^2$.
Then, we obtain
\begin{align}
	\{
		\textbf{Q}_{\tau,\rho},
		\textbf{Q}_{\tau',\rho'}
	\}
	&=
		\textbf{Q}_{2\tau'',2\rho''}
		- \left.\frac{1}{2}(D^zD_z^2 f')\right|_{z=w}
	.
\end{align}


\section{Discussion}\label{sec:discussion}

We have constructed standard and dual supertranslation charges on the future horizon of the Schwarzschild black hole using the first-order formalism of \cite{Godazgar:2020gqd,Godazgar:2020kqd}.
Then, we have explored the consequences of allowing for singularities in the parameter function of supertranslations.
Singular supertranslations arise naturally in the extended phase space associated with the BMS algebra \cite{Barnich:2011mi}.
Also, in electrodynamics singular large gauge transformations are closely related to Dirac string configurations in the bulk \cite{Freidel:2018fsk},
	and singular supertranslations can be considered as their gravitational analog.
Using a simple pole as an example, we have demonstrated that singularities lead to the presence of a central term in the Dirac bracket charge algebra,
	implying that the symmetry algebra becomes anomalous.
In order to remove such a term, we have introduced a gravitational Chern-Simons theory \cite{Witten:1989ip} with gauge group $SL(2,\C)$ on the horizon.
Being a topological theory, this theory is suitable to live on the horizon which is a null surface, and in addition does not contribute a stress-energy tensor which may perturb the gravitational field.
We have shown that the large gauge transformation of this boundary theory can be organized such that its charge algebra cancels the anomalous central term of the bulk gravity theory.

Some comments are in order.
In this paper, we have shown that an $SL(2,\C)$ Chern-Simons theory on the horizon can cancel the central term,
	but what we have not shown is that this theory is unique in being capable of this job.
Whether there exist other topological field theories that can cancel the central term is an interesting question, as the properties shared by the set of such theories
	will teach us more about the fundamental nature of the structure on the black hole horizon.

Since the standard and dual supertranslation algebra on the horizon is an asymptotic symmetry algebra and hence is not gauged, one may observe the anomalous central term
	and decide that we extend the symmetry algebra to incorporate such a term instead of removing it.
As an example of this viewpoint, central extension of classical asymptotic symmetry algebra is also present in the literature such as \cite{Brown:1986nw}.
It would be very interesting to explore this direction, as the work of Brown and Henneaux is intimately related to the existence of a dual two-dimensional holographic boundary CFT.
We leave this for future investigation.

In electromagnetism, there are specific examples of configurations that are associated with singular gauge transformations \cite{Freidel:2018fsk}.
Then, one may ask whether there are well-known gravitational configurations associated with singular supertranslations.
It has been shown by Strominger and Zhiboedov \cite{Strominger:2016wns} that finite superrotations at the null infinity
	map asymptotically flat spacetimes to spacetimes with isolated defects, which are interpreted as cosmic strings.
It is not clear whether singular supertranslations can have similar effects.
It would be very interesting to see find such an example associated with singular supertranslations.

Finally, the structure of null infinity is very similar to the future Schwarzschild horizon, and thus we expect a similar structure to be present at the null infinity as well.
It would be interesting to explore how such a structure could affect scattering amplitudes.

\acknowledgments
SC thanks the participants of the Corfu 2022 Workshop on Celestial Amplitudes and Flat Space Holography for stimulating discussions.
MJP acknowledges funding from the Science and Technology Facilities Council (STFC)
Consolidated Grant ST/T000686/1 ``Amplitudes, Strings and duality''. MJP would also like
to thank the UK STFC for financial support under grant ST/L000415/1. No new data were
generated or analysed during this study. The work of SC is supported by the European
Research Council (ERC) under the European Union's Horizon 2020 research and innovation
programme (grant agreement No 852386). SC also acknowledges financial support from the
Samsung Scholarship.

\appendix

\section{Modified Lie bracket} \label{app:modified}

In this appendix
we describe, in detail, the construction of the modified Lie bracket \cite{Barnich:2011mi}
on the Schwarzschild horizon ${\cal H}^+.$ 

The vector field $\xi$ that generates a supertranslation $f(\x)$ and a superrotation $Y(\x)$ is
\begin{align}
	\xi
	&=
		\left(
			f
			+ \frac{v}{2}\psi
		\right)\p_v
		- \frac{1}{2}\left(
			D^2(f + \frac{v}{2}\psi)
			+ r\psi
		\right)\p_r
		+ \left(
			\frac{1}{r}D^A(f + \frac{v}{2}\psi)
			+ Y^A
		\right)\p_A
		\label{ST+SR}
	,
\end{align}
where $\psi \equiv D_A Y^A$.
Let 
\begin{align}
	F(v,\x) \equiv f(\x) + \frac{v}{2}\psi(\x)
	,
\end{align}
such that $\p_v F = \frac{1}{2}\psi$.
Then,
\begin{align}
	\xi
	&=
		F\p_v
		- \frac{1}{2}D^2 F \p_r
		+ \frac{1}{r}D^A F \p_A
		- \frac{r}{2}\psi \p_r
		+ Y^A\p_A
	\nonumber,\\
	\xi_v
	&=
		-\Lambda F
		- \frac{1}{2}D^2 F
		- \frac{r}{2}\psi
	,\qquad
	\xi_r
	=
		F
	,\qquad
	\xi_A
	=
		rD_A F
		+r^2 Y_A
	,
\end{align}
where $Y_A = \gamma_{AB} Y^B$.
In this form, $\xi$ is like a $v$-dependent supertranslation $F$ but with ``corrections'' 
$-\frac{r}{2}\psi\p_r + Y^A\p_A$.
Since we are only interested in terms linear in $\xi$, we can compute the contributions of $F$ 
and the remainders separately.

For the unperturbed Schwarzschild spacetime, the non-vanishing Christoffel symbols $\bar\Gamma^a_{bc}$
are
\begin{gather}
	\bar\Gamma^v_{vv} = \frac{M}{r^2}
	,\quad
	\bar\Gamma^v_{AB} = -r\gamma_{AB}
	,\quad
	\bar\Gamma^r_{vv} = \frac{M\Lambda}{r^2}
	,\quad
	\bar\Gamma^r_{vr} = -\frac{M}{r^2}
	\nonumber,\\
	\bar\Gamma^r_{AB} = -r\Lambda\gamma_{AB}
	,\quad
	\bar\Gamma^A_{rB} = \frac{1}{r}\delta^A_B
	,\quad
	\bar\Gamma^A_{BC} = {}^{(2)}\Gamma^A_{BC}
	.
\end{gather}
Now, the metric perturbations $\bar\delta g_{ab}$ generated by $\xi$ are
 $\delta\bar g_{ab} \equiv \mL_\xi \bar g_{ab}$ and so we find
\begin{align}
	\delta\bar g_{vv}
	&=
		\frac{M}{r^2} D^2 F
		- \psi
		+ \frac{3M}{r} \psi
		- \frac{1}{2}D^2 \psi
	,\\
	\delta\bar g_{vr}
	&=
		0
	,\\
	\delta \bar g_{vA}
	&=
		- D_A\left(
			\Lambda F
			+ \frac{1}{2} D^2 F
		\right)
	,\\
	\delta \bar g_{AB}
	&=
		2r D_AD_B F
		- r\gamma_{AB} D^2 F
		+ r^2\left(
			D_AY_B
			+ D_BY_A
			- \gamma_{AB}\psi
		\right)
	.
\end{align}
The perturbed metric is then
\begin{align}
	ds^2
	&=
		-\left(
			\Lambda
			- \frac{M}{r^2}D^2 F
			+ \psi
			- \frac{3M}{r} \psi
			+ \frac{1}{2}D^2 \psi
		\right)dv^2
		+ 2dvdr
		- D_A\left(2\Lambda F + D^2 F\right) dvd\x^A
		\nonumber\\&\quad
		+ \left[
			r^2\gamma_{AB}
			+ 2rD_AD_B F
			- r\gamma_{AB} D^2 F
			+ r^2\left(
				D_AY_B
				+ D_BY_A
				- \gamma_{AB}\psi
			\right)
		\right]d\x^Ad\x^B
	.
\end{align}
Using this and the relation
\begin{align}
	\Gamma^a_{bc}
	&=
		\bar\Gamma^a_{bc}
		+ \frac{1}{2}\bar g^{ad}
		\left(
			\bar\nabla_b \delta\bar g_{dc}
			+ \bar\nabla_c \delta\bar g_{db}
			- \bar\nabla_d \delta\bar g_{bc}
		\right)
		+ \O(\delta\bar g^2)
	,
\end{align}
we compute some of the perturbed Christoffel symbols to linear order in $\xi$,
\begin{align}
	\Gamma^v_{rr}
	&=
	\Gamma^r_{rr}
	=
	\Gamma^A_{rr}
	= 0
	,\\
	\Gamma^v_{rA}
	&=
		0
	,\\
	\Gamma^r_{rA}
	&=
		\frac{1}{r} D_A F
		- \frac{3M}{r^2} D_A F
		+ \frac{1}{2r} D_AD^2 F
	,\\
	\Gamma^B_{rA}
	&=
		\frac{1}{r}\delta^B_A
		- \frac{1}{2r^2}
		\left(
			2D^BD_A F
			- \delta^B_AD^2 F
		\right)
	,
\end{align}
which turn out to be exactly the same as the components of supertranslated metric with just $f\to F$. 
Also
\begin{align}
	\gamma^{AB}\Gamma^v_{AB}
	&=
		-2r
	,\\
	\gamma^{AB}\Gamma^r_{AB}
	&=
		- 2r\Lambda
		- D^2 F
		+ \frac{4M}{r} D^2 F
		- 2r \psi
		+ 6M \psi
		- r D^2 \psi
		- \frac{1}{2} D^2 D^2 F
	,\\
	\gamma^{AB}\Gamma^C_{AB}
	&=
		\gamma^{AB}{}^{(2)}\Gamma^C_{AB}
		+ \frac{4M}{r^2} D^C F
		+ Y^C
		+ D^2 Y^C
	.
\end{align}
Using the above, we can write for any vector field $\zeta_a$
\begin{align}
	\nabla_r \zeta_r
	&=
		\p_r \zeta_r
	,\\
	\nabla_r \zeta_A + \nabla_A \zeta_r
	&=
		\p_r \zeta_A
		+ D_A \zeta_r
		- \zeta_r \left(
			\frac{2}{r} D_A F
			- \frac{6M}{r^2} D_A F
			+ \frac{1}{r} D_AD^2 F
		\right)
		\nonumber\\&\quad
		- \frac{2}{r}\zeta_A
		+ \frac{1}{r^2} \zeta_B \left(
			2D^BD_A F
			- \delta^B_AD^2 F
		\right)
	,\\
	\gamma^{AB}\nabla_A \zeta_B
	&=
		\zeta_r \left(
			D^2 F
			- \frac{4M}{r} D^2 F
			+ 2r \psi
			- 6M \psi
			+ r D^2 \psi
			+ \frac{1}{2} D^2 D^2 F
			+ 2r\Lambda
		\right)
		\nonumber\\&\quad
		+ D^A \zeta_A
		+ 2r \zeta_v
		- \zeta_C \Bigg(
			\frac{4M}{r^2} D^C F
			+ Y^C
			+ D^2 Y^C
		\Bigg)
	.
\end{align}
Now we can relate the components of any contravariant vector field $\zeta^a$
to the components of a covaraint vector field $\zeta_a$
using the perturbed metric,
\begin{align}
	\zeta_v
	&=
		- \Lambda \zeta^v
		+ \zeta^v\left(
			\frac{M}{r^2} D^2 F
			- \psi
			+ \frac{3M}{r} \psi
			- \frac{1}{2}D^2 \psi
		\right)
		+ \zeta^r
		- \zeta^A D_A\left(
			\Lambda F
			+ \frac{1}{2} D^2 F
		\right)
	,\\
	\zeta_r
	&=
		\zeta^v
	,\\
	\zeta_A
	&=
		- \zeta^v D_A\left(
			\Lambda F
			+ \frac{1}{2} D^2 F
		\right)
		+ r^2 \gamma_{AB} \zeta^B
		\nonumber\\&\quad
		+ \zeta^B \left(
			2r D_AD_B F
			- r\gamma_{AB} D^2 F
			+ r^2\left(
				D_AY_B
				+ D_BY_A
				- \gamma_{AB}\psi
			\right)
		\right)
	.
\end{align}
Now let $\zeta^a$ to be a new Schwarzschild supertranslation plus superrotation vector field 
parametrized by $g(\x)$ and $Z^A(\x)$.
	Employing the shorthand $\phi \equiv D_A Z^A$ and $G \equiv g + \frac{v}{2}\phi$,
\begin{align}
	\zeta^v = G + \delta \zeta^v
	,\qquad
	\zeta^r = -\frac{1}{2}D^2 G - \frac{r}{2}\phi + \delta \zeta^r
	,\qquad
	\zeta^A = \frac{1}{r}D^A G + Z^A + \delta\zeta^A
	,
\end{align}
where $\delta\zeta^a$ is the change in $\zeta^a$ due to the original diffeomorphism $\xi^a$.
To first order in the perturbation, 
\begin{align}
	\zeta_v
	&=
		- \Lambda \delta\zeta^v
		+ \delta\zeta^r
		+ G\left(
			- \Lambda
			+ \frac{M}{r^2} D^2 F
			- \psi
			+ \frac{3M}{r} \psi
			- \frac{1}{2}D^2 \psi
		\right)
		- \frac{1}{2}D^2 G
		- \frac{r}{2}\phi
		\nonumber\\&\quad
		- \left(
			\frac{1}{r} D^A G
			+ Z^A
		\right)
		\left(
			\Lambda D_A F
			+ \frac{1}{2} D_A D^2 F
		\right)
	,\\
	\zeta_r
	&=
		G + \delta\zeta^v
	,\\
	\zeta_A
	&=
		\left(
			\frac{1}{r}D^B G
			+ Z^B
		\right)
		\left(
			2r D_AD_B F
			- r\gamma_{AB} D^2 F
			+ r^2\left(
				D_AY_B
				+ D_BY_A
				- \gamma_{AB}\psi
			\right)
		\right)
		\nonumber\\&\quad
		- G \left(
			\Lambda D_A F
			+ \frac{1}{2} D_A D^2 F
		\right)
		+ r D_A G
		+ r^2 Z_A
		+ r^2 \gamma_{AB} \delta \zeta^B
	.
\end{align}
Plugging back in and demanding that $\nabla_r \zeta_r = \nabla_A \zeta_r+\nabla_r\zeta_A=\gamma^{AB}\nabla_A\zeta_B=0$, we obtain
\begin{align}
	0
	&=
		\p_r \delta\zeta^v
	,\\
	0
	&=
		r^2\gamma_{AB} \p_r\delta \zeta^B
		+ D_A \delta\zeta^v
		- \frac{2}{r} (D^B G) D_AD_B F
		\nonumber\\&\quad
		+ \frac{1}{r} (D_A G) D^2 F
		- (D^B G) (D_AY_B+D_BY_A)
		+ (D_A G) \psi
	,\\
	0
	&=
		- \Lambda(D^AG) D_A F
		+ 2(D^AD^B G) D_AD_B F
		+ \frac{r^2}{2} (D^AZ^B + D^BZ^A)(D_AY_B + D_BY_A)
		\nonumber\\&\quad
		- (D^2 G) D^2 F
		- r^2 \phi\psi
		- r (D^2 G)\psi
		- r (D^2 F)\phi
		+ 2r(D^AD^B G) D_AY_B
		\nonumber\\&\quad
		- \frac{1}{2} (D^AG) D_A D^2 F
		+ 2r (D^AZ^B) D_AD_B F
		+ r^2 D_A \delta \zeta^A
		+ 2r\delta\zeta^r
	.
\end{align}
Solving for $\delta\zeta$, we obtain
\begin{align}
	\delta\zeta^v
	&=
		0
	,\\
	\delta\zeta^r
	&=
		\frac{1}{2r}\left(
			\Lambda(D^AG) D_A F
			- (D^AD^B G) D_AD_B F
			+ \frac{1}{2} (D^2 G) D^2 F
		\right)
		- r D^{(A}Z^{B)}D_{(A}Y_{B)}
		\nonumber\\&\quad
		+ \frac{1}{2r}(D^AG) D^2 D_A F
		- (D^AZ^B) D_AD_B F
		+ \frac{1}{2} (D^2 F)\phi
		+ \frac{1}{2} (D_B G) D^2Y^B
		\nonumber\\&\quad
		+ \frac{1}{2} (D_B G) Y^B
		+ \frac{r}{2} \phi\psi
	,\\
	\delta\zeta^A
	&=
		- \frac{1}{r^2} (D^B G) D^AD_B F
		+ \frac{1}{2r^2} (D^A G) D^2 F
		- \frac{1}{r} (D_B G) (D^AY^B+D^BY^A)
		\nonumber\\&\quad
		+ \frac{1}{r} (D^A G) \psi
	.
\end{align}
We need to remind ourselves here that these $\delta\zeta^a$ are the changes in $\zeta^a$ 
due to the transformation $\xi^a$.
Due to this nature of $\delta\zeta^a$, we will change our notation to 
$\delta \zeta^a \to \delta_\xi \zeta^a$.
The changes in $\xi^a$ due to $\zeta^a$ can be obtained by exchanging $\xi\leftrightarrow \zeta$, and we will denote this as $\delta_\zeta \xi^a$.

The regular Lie bracket $[\xi,\zeta]^a = \xi^b\p_b\zeta^a-\zeta^b\p_b\xi^a$ of two vector fields can be computed straightforwardly from \eqref{ST+SR},
\begin{align}
	[\xi,\zeta]^v
	&=
		\frac{1}{2} F \phi
		- \frac{1}{2} G \psi
		+ Y^A D_A G
		- Z^A D_A F
	,\\
	[\xi,\zeta]^r
	&=
		- \frac{1}{4} F D^2 \phi
		+ \frac{1}{4} G D^2 \psi
		+ \frac{1}{4}\phi D^2 F
		- \frac{1}{4}\psi D^2 G
		- \frac{1}{2r}(D^A F)D_AD^2 G
		+ \frac{1}{2r}(D^A G)D_AD^2 F
		\nonumber\\&\quad
		- \frac{1}{2}(D^A F)D_A\phi
		+ \frac{1}{2}(D^A G)D_A\psi
		- \frac{1}{2}Y^A D_AD^2 G
		+ \frac{1}{2}Z^A D_AD^2 F
		- \frac{r}{2}Y^A D_A\phi
		\nonumber\\&\quad
		+ \frac{r}{2}Z^A D_A\psi
	,\\
	[\xi,\zeta]^A
	&=
		\frac{1}{2r} F D^A \phi
		- \frac{1}{2r} G D^A \psi
		+ \frac{1}{2r^2} (D^2 F) D^A G
		- \frac{1}{2r^2} (D^2 G) D^A F
		+ \frac{1}{2r} \psi D^A G
		- \frac{1}{2r} \phi D^A F
		\nonumber\\&\quad
		+ \frac{1}{r^2}(D^BF)D_B D^A G
		- \frac{1}{r^2} (D^BG)D_B D^A F
		+ \frac{1}{r}Y^BD_B D^A G
		- \frac{1}{r} Z^BD_B D^A F
		\nonumber\\&\quad
		+ \frac{1}{r}(D^B F)D_B Z^A
		- \frac{1}{r}(D^B G)D_B Y^A
		+ Y^BD_B Z^A
		- Z^BD_B Y^A
	.
\end{align}
We define the modified bracket by correcting this by $\delta_\xi\zeta^a$ and $\delta_\zeta\xi^a$,
\begin{align}
	[\xi,\zeta]^a_M
	&=
		[\xi,\zeta]^a
		- \delta_\xi \zeta^a
		+ \delta_\zeta \xi^a
	.
\end{align}
Using the expressions for $\delta_\xi\zeta^a$ that we have computed earlier, we obtain
\begin{align}
	[\xi,\zeta]^v_M
	&=
		\frac{1}{2} F \phi
		+ Y^A D_A G
		- (\xi\leftrightarrow \zeta)
	,\\
	[\xi,\zeta]^r_M
	&=
		- \frac{1}{4} F D^2 \phi
		- \frac{1}{4} (D^2 F)\phi
		- \frac{1}{2} (D^A F)D_A\phi
		- \frac{1}{2}Y^A D^2 D_A G
		- (D^AY^B) D_AD_B G
		\nonumber\\&\quad
		- \frac{1}{2} (D_B G) D^2Y^B
		- \frac{r}{2} Y^A D_A\phi
		- (\xi\leftrightarrow \zeta)
	,\\
	[\xi,\zeta]^A_M
	&=
		\frac{1}{2r} F D^A \phi
		- \frac{1}{2r} \psi D^A G
		+ \frac{1}{r} Y^B D_B D^A G
		+ Y^BD_B Z^A
		+ \frac{1}{r} (D_B G) D^AY^B
		\nonumber\\&\quad
		- (\xi\leftrightarrow \zeta)
	.
\end{align}
The $v$-component can be reorganized as
\begin{align}
	[\xi,\zeta]^v
	&=
		\frac{1}{2} f \phi
		- \frac{1}{2} g \psi
		+ Y^A D_A g
		- Z^A D_A f
		+ \frac{v}{2}D_A\left(
			Y^B D_B Z^A
			- Z^B D_B Y^A
		\right)
	.
\end{align}
Let us define
\begin{align}
	\hat f
	&=
		\frac{1}{2} f \phi
		- \frac{1}{2} g \psi
		+ Y^A D_A g
		- Z^A D_A f
	,\\
	\hat Y^A &=
		Y^B D_B Z^A
		- Z^B D_B Y^A
	.
\end{align}
Then, define $\hat \psi \equiv D_A \hat Y^A$, and take $\hat F \equiv \hat f + \frac{v}{2}\hat\psi$ so that we have $[\xi,\zeta]^v=\hat F$,
\begin{align}
	\hat F
	&=
		\frac{1}{2}F\phi + Y^AD_A G - (\xi\leftrightarrow\zeta)
	.
\end{align}
With this definition, observe that we have exactly the modified bracket components
\begin{align}
	- \frac{1}{2}D^2 \hat F
	- \frac{r}{2}\hat \psi
	&=
		- \frac{1}{4} F D^2\phi
		- \frac{1}{4} (D^2F)\phi
		- \frac{1}{2} (D^AF) D_A\phi
		- \frac{1}{2} Y^A D^2 D_A G
		\nonumber\\&\quad
		- (D^AY^B)D_AD_B G
		- \frac{1}{2} (D_A G)D^2Y^A
		- \frac{r}{2}Y^AD_A \phi
		- (\xi\leftrightarrow \zeta)
	\\ &=
		[\xi,\zeta]^r_M
	,
\end{align}
and
\begin{align}
	\frac{1}{r}D^A \hat F
	+ \hat Y^A
	&=
		\frac{1}{2r} FD^A \phi
		- \frac{1}{2r}\psi D^A G
		+ \frac{1}{r}Y^B D_BD^A G
		+ Y^B D_BZ^A
		+ \frac{1}{r}(D_B G)D^AY^B
		\nonumber\\&\quad
		- (\xi\leftrightarrow \zeta)
	\\ &=
		[\xi,\zeta]^A_M
	.
\end{align}
This implies that
\begin{align}
	[\xi,\zeta]_M
	&=
		\left(\hat f + \frac{v}{2}\hat \psi\right)\p_v
		- \frac{1}{2}\left(
			D^2 \left(\hat f + \frac{v}{2}\hat \psi\right)
			+ r\hat \psi
		\right)\p_r
		+ \left(
			\frac{1}{r}D^A \left(\hat f + \frac{v}{2}\hat \psi\right)
			+ \hat Y^A
		\right)\p_A
	.
\end{align}
Comparing the RHS to the expression \eqref{ST+SR}, we can see that it is another supertranslation $\hat f$ together with superrotation $\hat Y^A$.

We conclude that given two pairs $(f_1, Y_1), (f_2,Y_2)$ of supertranslation and superrotation, the modified bracket has the algebra
\begin{align}
	[(f_1,Y_1),(f_2,Y_2)]_M = (\hat f,\hat Y),
\end{align}
with the product being another supertranslation together with a superrotation parametrized by
\begin{align}
	\hat f
	&=
		\frac{1}{2} f_1 D_AY^A_2
		- \frac{1}{2} f_2 D_AY^A_1
		+ Y_1^A D_A f_2
		- Y_2^A D_A f_1
	,\\
	\hat Y^A &=
		Y_1^B D_B Y_2^A
		- Y_2^B D_B Y_1^A
	,
\end{align}
which is equivalent to the BMS algebra at the null infinity \cite{Barnich:2011mi}.


\section{Derivation of horizon charges} \label{app:charges}

In this section, we will give a derivation of the supertranslation and dual supertranslation charges using the formula of \cite{Godazgar:2020gqd,Godazgar:2020kqd},
\begin{align}
	\slashed \delta Q_E^\H
	&=
		\frac{1}{16\pi} \ep_{\abcd}
			\int_{\p \mH^+}
			(i_\xi E^\c)\delta\w^{\ab}\wedge E^\d
		\label{QEa}
	,\\
	\slashed \delta Q_M^\H
	&=
		\frac{1}{8\pi}\int_{\p \mH^+}
			(i_\xi E^\alpha)\delta\w_\ab \wedge E^\beta
		\label{QMa}
	.
\end{align}
Here $\w_\ab$ is the (torsion-free) spin connection 1-form, and $\delta \w$ is the change in $\w$ induced by the variation $\delta g_{ab} = h_{ab}$ of the metric.

In order to incorporate the variation of the metric, we will parametrize a generic metric in Bondi gauge by
\begin{align}
	g_{ab}
	&=
		\begin{pmatrix}
			V + W_A W^A & U & W_B \\
			U & 0 & 0 \\
			W_A & 0 & g_{AB}
		\end{pmatrix}
		,
\end{align}
where $V$, $U$, $W_A$ are real functions of $v$, $r$, $\x^A$.
The inverse metric is
\begin{align}
	g^{ab}
	&=
		\begin{pmatrix}
			0 & U^{-1} & 0 \\
			U^{-1} & -VU^{-2} & -U^{-1}W^B \\
			0 & -U^{-1}W^A & g^{AB}
		\end{pmatrix}
	,
\end{align}
where $g^{AB}$ is the inverse of the two-dimensional metric $g_{AB}$, and $W^A = g^{AB}W_B$ (not $\gamma^{AB}W_B$).
Since this metric may deviate from that of Schwarzschild,
	the two-dimensional curved indices $A,B,C,\ldots$ in this section, and only in this section,
	are raised and lowered using $g^{AB}$ and $g_{AB}$  rather than $\gamma^{AB}$ and
	$\gamma_{AB}$, the metric on the unit 2-sphere metric.

We will employ the following set of vielbein $E^\a = E^\a{}_a dx^a$,
\begin{align}
	E^1
	&=
		\frac{V}{2} dv + Udr
	,\\
	E^2
	&=
		-dv
	,\\
	E^3
	&=
		W_A \mu^A dv + \mu_A d\x^A
	,\\
	E^4
	&=
		W_A \bar\mu^A dv + \bar\mu_A d\x^A
	,
\end{align}
where $\mu_A$, $\bar\mu_A$ are complex functions of $v$, $r$, $\x^A$,
	and $\mu^A=g^{AB}\mu_B$, $\bar\mu^A = g^{AB} \bar\mu_B$
(bar denotes complex conjugation, so $\bar\mu_A$ is the complex conjugate of $\mu_A$ and hence $E^3 = \overline{E^4}$).
They satisfy the conditions
\begin{align}
	\mu_A\bar\mu_B + \mu_B\bar\mu_A &= g_{AB}
	,\qquad
	\mu^A \bar\mu_A = 1
	,\qquad
	\mu^A \mu_A = \bar\mu^A \bar \mu_A = 0
	.
\end{align}
The tangent space metric and its inverse are
\begin{align}
	\eta_\ab = \eta^\ab = \begin{pmatrix}
		0&-1&0&0\\
		-1&0&0&0\\
		0&0&0&1\\
		0&0&1&0
	\end{pmatrix}
	,
\end{align}
and the inverse vielbeins $E_\alpha = E_\alpha{}^a\p_a$ are
\begin{align}
	E_1
	&=
		U^{-1}\p_r
	,\\
	E_2
	&=
		-\p_v + \frac{V}{2U} \p_r + W^A \p_A
	,\\
	E_3
	&=
		\bar\mu^A\p_A
	,\\
	E_4
	&=
		\mu^A \p_A
	.
\end{align}
One can readily check that
\begin{align}
	E_\alpha{}^a E^\alpha{}_b
	=
		\delta^a{}_b
	&,\qquad
	E_\alpha{}^a E^\beta{}_a
	=
		\delta_\alpha{}^\beta
	,\\
	E^\alpha{}_a E^\beta{}_b \eta_\ab = g_{ab}
	&,\qquad
	E_\alpha{}^a E_\beta{}^b \eta^\ab = g^{ab}
	.
\end{align}
The spin connection 1-form $\w_\ab$ is defined as
\begin{align}
	dE^\alpha &= -\w^\alpha{}_\beta\wedge E^\beta = \frac{1}{2}c^\alpha{}_{\beta\gamma}E^\beta \wedge E^\gamma
	,\\
	\w_\ab &= \frac{1}{2}(c_{\alpha\beta\gamma} - c_{\beta\alpha\gamma} - c_{\gamma\ab})E^\gamma
	,
\end{align}
where $c_{\a\b\c}$ are the anholonomy coefficients.
Explicit expressions for the coefficients read
\begin{align}
	c_{1\b\c}
	&=
	0
	,\\
	c_{212}
	&=
		\frac{1}{U}\left(\frac{1}{2}V'-\dot U + W^A\p_A U\right)
	,\\
	c_{213}
	&=
		\frac{\bar\mu^A\p_A U}{U}
	,\\
	c_{223}
	&=
		- \frac{1}{2}\bar\mu^A\p_A V
		+ \frac{\bar\mu^A\p_A U}{2U}V
	,\\
	c_{234}
	&=
		0
	,\\
	c_{312}
	&=
		-\frac{W^A{}'\bar\mu_A}{U}
	,\\
	c_{313}
	&=
		\frac{\bar\mu^A\bar\mu_A'}{U}
	,\\
	c_{314}
	&=
		\frac{\mu^A\bar\mu_A'}{U}
	,\\
	c_{323}
	&=
		\bar\mu^A \p_A (W\cdot \bar\mu)
		- \bar\mu^A \dot {\bar\mu}_A
		+ \frac{\bar\mu^A\bar\mu_A'}{2U}V
		+ W^A \bar\mu^B (\p_A \bar\mu_B-\p_B \bar\mu_A)
	,\\
	c_{324}
	&=
		\mu^A \p_A (W\cdot \bar\mu)
		- \mu^A \dot {\bar\mu}_A
		+ \frac{\mu^A\bar\mu_A'}{2U}V
		+ W^A \mu^B (\p_A \bar\mu_B-\p_B \bar\mu_A)
	,\\
	c_{334}
	&=
		\left (\mu^A \bar\mu^B - \bar\mu^A \mu^B\right ) \p_B \bar\mu_A
	.
\end{align}
The remaining coefficients can be obtained using the antisymmetry $c_{\a\b\c}=-c_{\a\c\b}$ and the fact $E^3=\overline{E^4}$ implies switching indices $3\leftrightarrow 4$
	corresponds to complex conjugation, for instance $c_{213}=\overline{c_{214}}$ and $c_{434}=\overline{c_{343}}=-\overline{c_{334}}$.
Using this to compute $\w_{\ab}$, we obtain
\begin{align}
	\w_{12}
	&=
		\frac{1}{U}\left(-\frac{1}{2}V'+\dot U-W^A\p_A U\right) E^2
		\nonumber\\&\quad
		+ \frac{1}{2U}\left(
			- \bar\mu^A\p_A U
			+ {W^A}'\bar\mu_A
		\right) E^3
		+ \frac{1}{2U}\left(
			- \mu^A\p_A U
			+ {W^A}'\mu_A
		\right) E^4
	,\\
	\w_{13}
	&=
		\frac{1}{2U}\left(
			W^A{}'\bar\mu_A
			- \bar\mu^A \p_A U
		\right) E^2
		- \frac{\bar\mu^A \bar\mu_A'}{U}E^3
		- \frac{1}{2U}\left(
			\mu^A \bar\mu_A'
			+ \bar\mu^A \mu_A'
		\right)E^4
	,\\
	\w_{23}
	&=
		\frac{1}{2U}\left(
			-\bar\mu^A\p_A U
			- W^A{}'\bar\mu_A
		\right)E^1
		+ \frac{1}{2}\left(
			\bar\mu^A\p_A V
			- \frac{\bar\mu^A\p_A U}{U}V
		\right)E^2
		\nonumber\\&\quad
		- \left(
			\bar\mu^A \p_A (W\cdot \bar\mu)
			- \bar\mu^A \dot {\bar\mu}_A
			+ \frac{\bar\mu^A\bar\mu_A'}{2U}V
			+ W^A \bar\mu^B (\p_A \bar\mu_B-\p_B \bar\mu_A)
		\right)E^3
		\nonumber\\&\quad
		- \frac{1}{2}\left(
			\mu^A \p_A (W\cdot \bar\mu)
			- \mu^A \dot {\bar\mu}_A
			+ \frac{\mu^A\bar\mu_A'}{2U}V
			+ W^A \mu^B (\p_A \bar\mu_B-\p_B \bar\mu_A)
			+ \cc
		\right)E^4
		\label{w}
	,\\
	\w_{34}
	&=
		\frac{1}{2U}\left(
			- \mu^A\bar\mu_A'
			+ \bar\mu^A\mu_A'
		\right)E^1
		\nonumber\\&\quad
		+ \frac{1}{2}\left(
			\bar\mu^A\p_A (W\cdot\mu)
			- \bar\mu^A\dot\mu_A
			+ \frac{\bar\mu^A\mu_A'}{2U}V
			+ W^A \bar\mu^B (\p_A \mu_B - \p_B \mu_A )
			-\cc
		\right)E^2
		\nonumber\\&\quad
		- \left (\mu^A \bar\mu^B - \bar\mu^A \mu^B\right ) \p_B \bar\mu_A E^3
		- \left(
			\mu^A \bar\mu^B
			- \bar\mu^A \mu^B
		\right) \p_B \mu_A E^4
	.
\end{align}
We keep in mind that $E^3=\overline{E^4}$.
The remaining components can be obtained by antisymmetry and complex conjugation, for instance $\w_{42}=\overline{\w_{32}}=-\overline{\w_{23}}$.

\subsection{Supertranslation charge}
The conserved electric charge involves the differential form
\begin{align}
		\frac{1}{16\pi} \ep_{\abcd}
			(i_\xi E^\c)\delta\w^{\ab}\wedge E^\d
	.
\end{align}
We are interested in integrating
\begin{align}
	\frac{1}{2}\ep_{\abcd}(i_\xi E^\c) \delta \w^{\ab} \wedge E^\d
	&=
		\ep_{1234}i_\xi E^3 \delta \w^{12} \wedge E^4
		+ \ep_{1324}i_\xi E^2 \delta \w^{13} \wedge E^4
		+ \ep_{2314}i_\xi E^1 \delta \w^{23} \wedge E^4
		\nonumber\\&\quad
		+ \ep_{1243}i_\xi E^4 \delta \w^{12} \wedge E^3
		+ \ep_{1423}i_\xi E^2 \delta \w^{14} \wedge E^3
		+ \ep_{2413}i_\xi E^1 \delta \w^{24} \wedge E^3
		\nonumber\\&\quad
		+\cdots
\end{align}
over $S^2$.
Observe that the alternating tensor $\ep_\abcd$ is purely imaginary,
\begin{align}
	\overline{\ep_{1234}} = \ep_{1243} = -\ep_{1234}.
\end{align}
By explicit computation, one finds that
\begin{align}
	\ep_{1234} = E_1{}^aE_2{}^bE_3{}^cE_4{}^d \ep_{abcd} = -i,
\end{align}
where $\ep_{abcd}$ is the alternating tensor in the curved coordinates with $\ep_{vr\t\phi} = \sqrt{-\det g} = U r^2\sin\t$.
Using this and rearranging the indices, we obtain
\begin{align}
	\frac{i}{2}\ep_{\abcd}(i_\xi E^\c) \delta \w^{\ab} \wedge E^\d
	&=
		- i_\xi E^3 \delta \w_{12} \wedge E^4
		+ i_\xi E^2 \delta \w_{24} \wedge E^4
		- i_\xi E^1 \delta \w_{14} \wedge E^4
		\nonumber\\&\quad
		+ i_\xi E^4 \delta \w_{12} \wedge E^3
		- i_\xi E^2 \delta \w_{23} \wedge E^3
		+ i_\xi E^1 \delta \w_{13} \wedge E^3
		\nonumber\\&\quad
		+\cdots
	.
\end{align}
Let us look at this expression term by term.
We are interested only in coefficients of $E^3\wedge E^4$ as we are integrating a the two-sphere
on the horizon.
The first and fourth terms combine to yield
\begin{align}
	- i_\xi E^3 \delta \w_{12} \wedge E^4
	+ i_\xi E^4 \delta \w_{12} \wedge E^3
	&=
		\frac{1}{2}\xi^A
		\left(
			\p_A h_{vr}
			+ \frac{2}{r}h_{vA}
			- \p_rh_{vA}
		\right)
		E^3\wedge E^4
		\nonumber\\&\quad
		+ \cdots
	.
\end{align}
For the second term we have
\begin{align}
	i_\xi E^2 \delta \w_{24} \wedge E^4
	&=
		\frac{\xi^v}{2}\left(
			\frac{1}{r}h_{vv}
			+ \p_A h^A{}_v
			+ h^A{}_v \left(
				\bar\mu^B \p_A \mu_B
				+ \mu^B \p_A \bar\mu_B
			\right)
		\right)E^3\wedge E^4
		\nonumber\\&\quad
		+\cdots
	,
\end{align}
where we have used $\delta(\bar\mu^A\dot\mu_A) = \delta(\mu^A\dot{\bar\mu}_A)=0$.
It turns out that
\begin{align}
	\p_A h^A{}_v
	+ h^A{}_v \left(
		\bar\mu^B \p_A \mu_B
		+ \mu^B \p_A \bar\mu_B
	\right)
	=
	g^{AB} D_A h_{vB}
	= \frac{1}{r^2} \gamma^{AB}D_A h_{vB}
	,
\end{align}
where $D_A$ denotes covariant derivative on the unit 2-sphere (that is, compatible with $\gamma_{AB}$, not $g_{AB}$).
Thus, we can write
\begin{align}	
	i_\xi E^2 \delta \w_{24} \wedge E^4
	=
		\frac{\xi^v}{2}\left(
			\frac{1}{r}h_{vv}
			+ \frac{1}{r^2}\gamma^{AB}D_A h_{vB}
		\right)E^3\wedge E^4
		+\cdots
	.
\end{align}
The coefficient of $E^3\wedge E^4$ is real,
\begin{align}
	i_\xi E^2 \delta \w_{24} \wedge E^4
	- i_\xi E^2 \delta \w_{23} \wedge E^3
	&=
		\xi^v\left(
			\frac{1}{r}h_{vv}
			+ \frac{1}{r^2}\gamma^{AB}D_A h_{vB}
		\right)
		E^3\wedge E^4
		+\cdots		
	.
\end{align}
We also have
\begin{align}
	- i_\xi E^1 \delta \w_{14} \wedge E^4
	&=
		\xi^r \delta\left[
		\frac{1}{2U}\left(
			\bar\mu^A\mu_A'
			+ \mu^A\bar\mu_A'
		\right)\right]E^3
		\wedge E^4
		+ \frac{\xi^r}{2}\left(
			\bar\mu^A\mu_A'
			+ \mu^A\bar\mu_A'
		\right)\delta E^3
		\wedge E^4
	\nonumber\\&\quad
	+\cdots
	,\\
	i_\xi E^1 \delta \w_{13} \wedge E^3
	&=
		\xi^r \delta\left[
		\frac{1}{2U}\left(
			\mu^A \bar\mu_A'
			+ \bar\mu^A \mu_A'
		\right)\right]E^3\wedge E^4
		+
		\frac{\xi^r}{2}\left(
			\mu^A \bar\mu_A'
			+ \bar\mu^A \mu_A'
		\right)E^3\wedge \delta E^4
	\nonumber\\&\quad
	+\cdots
	.
\end{align}
Together we have
\begin{align}
	- i_\xi E^1 \delta \w_{14} \wedge E^4
	+ i_\xi E^1 \delta \w_{13} \wedge E^3
	&=
		-\frac{2\xi^r}{r} h_{vr} E^3\wedge E^4
		+ \frac{\xi^r}{r} \delta( E^3\wedge E^4)
		+ \cdots
	,
\end{align}
where we have used $\delta(\mu^A\bar\mu_A')=\delta(\bar\mu^A\mu_A')=0$.
With $\delta r=0$, we also have $\delta(E^3\wedge E^4)=0$ due to the Bondi gauge condition $\gamma^{AB}h_{AB}=0$.

Collecting the results, we obtain
\begin{align}
	\frac{i}{2}\ep_{\abcd}(i_\xi E^\c) \delta \w^{\ab} \wedge E^\d
	&=
		\Bigg[
			\frac{1}{2}\xi^A
			\left(
				\p_A h_{vr}
				+ \frac{2}{r}h_{vA}
				- \p_rh_{vA}
			\right)
			+ \xi^v\left(
				\frac{1}{r}h_{vv}
				+ \frac{1}{r^2}\gamma^{AB}D_A h_{vB}
			\right)
		\nonumber\\&\qquad
			-\frac{2\xi^r}{r} h_{vr}
		\Bigg]
		E^3\wedge E^4
		+\cdots
		.
\end{align}
Plugging this into \eqref{QEa}, we obtain the electric diffeomorphism charge associated with vector field $\xi$ on the Schwarzschild horizon $r=2M$ to be
\begin{align}
	\slashed \delta Q_E^\H
	&=
		\frac{M^2}{4\pi}\int d^2\x \sqrt\gamma
		\Bigg[
			\xi^A
			\left(
				\p_A h_{vr}
				+ \frac{1}{M}h_{vA}
				- \p_rh_{vA}
			\right)
		\nonumber\\&\hspace{30mm}
			+ \frac{1}{M}\xi^v\left(
				h_{vv}
				+ \frac{1}{2M}\gamma^{AB}D^A h_{vB}
			\right)
			-\frac{2\xi^r}{M} h_{vr}
		\Bigg]
		\label{aQE}
	.
\end{align}
For a smooth function $f(\Theta)$ and the horizon supertranslation vector field \eqref{STxi}, this formula is in exact agreement with the horizon supertranslation charge
	derived in \cite{Hawking:2016sgy}, as anticipated.

\subsection{Dual supertranslation charge}

The magnetic diffeomorphism charge associated with a vector field $\xi$ takes the form
\begin{align}
	\slashed\delta Q_M^\H
	&=
		\frac{i}{8\pi}\int_{\p\H}
			(i_\xi E^\alpha)\delta\w_\ab \wedge E^\beta
	.
\end{align}
Again, we only need to compute the $E^3 \wedge E^4$ component of the two-form
\begin{align}
	(i_\xi E^\alpha)\delta\w_\ab \wedge E^\beta
	.
\end{align}
The only part of the expression relevant to the $S^2$ integral is
\begin{align}
	(i_\xi E^\alpha)\delta\w_\ab \wedge E^\beta
	&=
		(i_\xi E^1)(\delta\w_{13} \wedge E^3+\delta\w_{14} \wedge E^4)
		+ (i_\xi E^2)(\delta\w_{23} \wedge E^3+\delta\w_{24} \wedge E^4)
		\nonumber\\&\quad
		+ (i_\xi E^3)\delta\w_{34} \wedge E^4
		+ (i_\xi E^4)\delta\w_{43} \wedge E^3
		+ \cdots
	,
\end{align}
where $\cdots$ contains all the irrelevant components.
Using the expression \eqref{w} for the spin connection, we can write
\begin{align}
	(\delta\w_{13} \wedge E^3+\delta\w_{14} \wedge E^4)|_{d\x^A\wedge d\x^B}
	&=
		- \delta\left[
			\frac{1}{2U}\left(
				\bar\mu^A\mu_A'
				+ \mu^A\bar\mu_A'
			\right)
		\right](E^3\wedge E^4+E^4\wedge E^3)
		\nonumber\\ &\quad
		- \frac{1}{2}\left(
			\bar\mu^A\mu_A'
			+ \mu^A\bar\mu_A'
		\right)(\delta E^3\wedge E^4+\delta E^4\wedge E^3)
		\nonumber\\ &\quad
		- (\bar\mu^A \bar\mu_A')\delta E^3\wedge E^3
		- (\mu^A\mu_A')\delta E^4\wedge E^4		
	.
\end{align}
The first line on the RHS is clearly zero since $E^3\wedge E^4+E^4\wedge E^3=0$.
The third line is also zero since
\begin{align}
	\bar\mu^A \bar\mu_A' = \frac{1}{r} \bar\mu^A \bar\mu_A = 0
	,\qquad
	\mu^A\mu_A' = \frac{1}{r}\mu^A\mu_A = 0
	.
\end{align}
In the second line, we have
\begin{align}
	\delta E^3\wedge E^4+\delta E^4\wedge E^3
	&=
		(\delta \mu_A \bar\mu_B + \delta \bar\mu_A \mu_B)d\x^A\wedge d\x^B
		\label{id1}
	.
\end{align}
One can show that the expression in parentheses on the RHS is $\frac{1}{2}h_{AB}$ and is therefore symmetric,
\begin{align}
	h_{AB} = \delta(\mu_A \bar\mu_B + \bar\mu_A \mu_B)
	= 2(\delta \mu_A \bar\mu_B + \delta \bar\mu_A \mu_B)
	,
\end{align}
which implies $\delta E^3\wedge E^4+\delta E^4\wedge E^3=0$.
Therefore we have
\begin{align}
	(\delta\w_{13} \wedge E^3+\delta\w_{14} \wedge E^4)|_{d\x^A\wedge d\x^B}
	=0.
\end{align}
The expression for $\delta\w_{23} \wedge E^3+\delta\w_{24} \wedge E^4$ is similar but with just more complicated coefficients.
To see this, first observe that the $E^3$ and $E^4$ components of $\w_{23}$ and $\w_{24}$ have the form
\begin{align}
	\w_{23}
	&=
		\cdots
		- A E^3
		- B E^4
	,\qquad
	\w_{24}
	=
		\cdots
		- BE^3
		- \bar AE^4
	,
\end{align}
where $A$ is complex and $B$ is real,
\begin{align}
	A
	&=
		\bar\mu^A \p_A (W\cdot \bar\mu)
		- \bar\mu^A \dot {\bar\mu}_A
		+ \frac{\bar\mu^A\bar\mu_A'}{2U}(V-W^2)
		+ W^A \bar\mu^B (\p_A \bar\mu_B-\p_B \bar\mu_A)
	,\\
	B
	&=
		\frac{1}{2}\left(
			\mu^A \p_A (W\cdot \bar\mu)
			- \mu^A \dot {\bar\mu}_A
			+ \frac{\mu^A\bar\mu_A'}{2U}(V-W^2)
			+ W^A \mu^B (\p_A \bar\mu_B-\p_B \bar\mu_A)
			+ \cc
		\right)
	.
\end{align}
Note that $A=B=0$ on Schwarzschild; it is only the variations $\delta A$ and $\delta B$ that do not necessarily vanish.
Thus, we have
\begin{align}
	(\delta\w_{23} \wedge E^3+\delta\w_{24} \wedge E^4)|_{d\x^A\wedge d\x^B}
	&=
		- (\delta B)(E^3\wedge E^4+E^4\wedge E^3)
		- B(\delta E^3\wedge E^4+\delta E^4\wedge E^3)
		\nonumber\\ &\quad
		- A\delta E^3\wedge E^3
		- \bar A\delta E^4\wedge E^4		
	\nonumber\\ &= 0
	,
\end{align}
where the second line vanishes since $A=B=0$, and the first line vanishes due to $E^3\wedge E^4+E^4\wedge E^3=0$.

At this point we are left with the two terms,
\begin{align}
	(i_\xi E^3)\delta\w_{34} \wedge E^4
	+ (i_\xi E^4)\delta\w_{43} \wedge E^3
	.
\end{align}
We first note that the $E^3$ and $E^4$ components of $\w_{34}=-\w_{43}$ can be written compactly using $\mu^A\bar\mu^B-\bar\mu^A\mu^B=i\ep^{AB}$ as
\begin{align}
	\w_{34}
	&=
		\cdots
		+ i\ep^{AB}\left(
			\p_A \bar\mu_B E^3
			+ \p_A \mu_B E^4
		\right)
		.
\end{align}
The variation $\delta \ep^{AB}$ is proportional to the trace $\gamma^{AB}h_{AB}$ and therefore vanishes in Bondi gauge.
Therefore if we vary $\w_{34}$, the variation only acts on the expression inside the parentheses,
\begin{align}
	\delta \w_{34}
	&=
		\cdots
		+ i\ep^{AB}\delta \left(
			\p_A \bar\mu_B E^3
			+ \p_A \mu_B E^4
		\right)
	\nonumber\\ &=
		\cdots
		+ i\ep^{AB}\left(
			\p_A \delta \bar\mu_B E^3
			+ \p_A \delta \mu_B E^4
			+ \p_A \bar\mu_B \delta E^3
			+ \p_A \mu_B \delta E^4
		\right)
		.
\end{align}
Plugging this in and using $i_\xi E^3=\xi^A\mu_A$ and $i_\xi E^4=\xi^A\bar\mu_A$, we obtain
\begin{align}
	(i_\xi E^3)\delta\w_{34} \wedge E^4
	+ (i_\xi E^4)\delta\w_{43} \wedge E^3
	&=
		i\ep^{AB}\xi^C\mu_C
		\left(
			\p_A \delta \bar\mu_B E^3
			+ \p_A \bar\mu_B \delta E^3
			+ \p_A \mu_B \delta E^4
		\right)\wedge E^4
		\nonumber\\&\quad
		- i\ep^{AB}\xi^C\bar\mu_C
		\left(
			\p_A \delta \mu_B E^4
			+ \p_A \bar\mu_B \delta E^3
			+ \p_A \mu_B \delta E^4
		\right)\wedge E^3
	\nonumber\\ &=
		\xi^C X_C
	,
\end{align}
where $X_C$ takes the form
\begin{align}
	X_C
	&=
		i\ep^{AB}\mu_C
		\left(
			\p_A \delta \bar\mu_B E^3
			+ \p_A \bar\mu_B \delta E^3
			+ \p_A \mu_B \delta E^4
		\right)\wedge E^4
		\nonumber\\&\quad
		- i\ep^{AB}\bar\mu_C
		\left(
			\p_A \delta \mu_B E^4
			+ \p_A \bar\mu_B \delta E^3
			+ \p_A \mu_B \delta E^4
		\right)\wedge E^3
	\nonumber\\ &=
		i\ep^{AB}\Big[
			\mu_C (\p_A \delta \bar\mu_B) \mu_D\bar\mu_E
			+ \mu_C (\p_A \bar\mu_B) \delta\mu_D\bar\mu_E
			+ \mu_C (\p_A \mu_B) \delta\bar\mu_D\bar\mu_E
			\nonumber\\&\quad
			+ \bar\mu_C (\p_A \delta \mu_B) \mu_D\bar\mu_E
			+ \bar\mu_C (\p_A \bar\mu_B) \mu_D\delta\mu_E
			+ \bar\mu_C (\p_A \mu_B) \mu_D\delta\bar\mu_E
		\Big]d\x^D\wedge d\x^E
	.
\end{align}
One finds that this expression is
\begin{align}
	X_C
	&=
		\frac{1}{2}
		\left(
			\p_\t\frac{h_{\phi\t}}{\sin\t}+\frac{2\cos\t}{\sin^2\t}h_{\phi\t}-\frac{\p_\phi h_{\t\t}}{\sin\t}
			,
			\sin\t\p_\t\frac{h_{\phi\phi}}{\sin^2\t}+\frac{2\cos\t}{\sin^2\t}h_{\phi\phi}-\p_\phi\frac{h_{\t\phi}}{\sin\t}
		\right)
		d\Omega
	\nonumber\\&=
		-\frac{r^2}{2}\ep^{AB}D_Ah_{BC} d\Omega
	,
	\label{XC}
\end{align}
where $d\Omega = \sin\t d\t\wedge d\phi$,
	and $D_A$ denotes the unit 2-sphere covariant derivative compatible with $\gamma_{AB}$.
Notice that $\ep^{AB}$ here is the Levi-Civita tensor for the metric $g_{AB}$, which contains the $r^2$ factor.
If we write $\bar\ep^{AB}$ for the Levi-Civita tensor corresponding to the $S^2$ metric $\gamma_{AB}$, we have the relation $\bar\ep^{AB} = r^2 \ep^{AB}$ and
\begin{align}
	X_C = -\frac{1}{2}\bar\ep^{AB} D_A h_{BC} d\Omega.
\end{align}
Collecting the results, we obtain the magnetic diffeomorphism charge associated with a vector field $\xi$ to be
\begin{align}
	\slashed\delta Q_M^\H
	&=
		\frac{1}{8\pi}\int_{\p\H}
			\xi^C X_C
	\nonumber\\ &=
		-\frac{1}{16\pi}\int_{\p\H}d^2\x\sqrt\gamma\,
			\xi^C \bar\ep^{AB}D_Ah_{BC}
	.
	\label{aQM}
\end{align}

\section{Dirac bracket of non-integrable piece} \label{app:nQE}

We can re-write $\nQE$ in terms of the delta function $\p_\zb f = 2\pi\delta^2(z-w)$.
Doing so and taking note that the covariant derivative $D_z$ is acting on a scalar and is therefore a plain partial derivative, we obtain
\begin{align}
	\nQE
	&=
		- \frac{1}{8\pi M}\int_\H dv\,d^2z\,(\p_\zb f) \p_z [D^2-1]^{-1}D^BD^A \sigma_{AB}
\end{align}
Partial integration in the second term by $\zb$ yields
\begin{align}
	\nQE
	&=
		\frac{1}{8\pi M}\int_\H dv\,d^2z\,(\p_z \p_\zb f) [D^2-1]^{-1}D^BD^A \sigma_{AB}
	.
\end{align}
The boundary term arising from this vanishes, since $\p_\zb f = 2\pi\delta^2(z-w)$ and the contour does not cross $w$.
To treat $[D^2-1]^{-1}$ explicitly, let us consider its Green's function $\GF(z,z')$ of $D^2-1$,\footnote{
	The Green's function depends on both $(z,\zb)$ and $(z',\zb')$, so we should have written $\GF(z,\zb,z',\zb')$ to be precise.
	We use the shorthand $\GF(z,z')$ for notational brevity.
}
\begin{align}
	(D^2-1)\GF(z,z') = \frac{1}{\gzz}\delta^2(z-z'),
\end{align}
which is derived in appendix \ref{app:green} to be,
\begin{align}
	\GF(z,z') = \frac{1}{4\sin(\pi \lambda)}P_{\lambda}(-\V n_z\cdot \V n_{z'})
	\label{Green}
	,
\end{align}
where $\lambda = \frac{1}{2}(-1 + i\sqrt 3)$, $P_\lambda$ is the Legendre function, and
\begin{align}
	\V n_z = \left(\frac{z+\zb}{1+z\zb},\frac{i(\zb-z)}{1+z\zb},\frac{1-z\zb}{1+z\zb}\right)
	\label{vector}
\end{align}
is the Cartesian coordinates of a unit vector on the sphere characterized by $(z,\zb)$.
The quantity $\V n_z\cdot \V n_{z'}$ reduces to $\cos\t$ when $(z',\zb')$ is set to the north pole, as it should.
Using $\GF$, we can write \eqref{nQE} as
\begin{align}
	\nQE
	&=
		\frac{1}{8\pi M}
			\int_\H dv\,d^2z\, (\p_z\p_\zb f) \int d^2z'\sqrt{\gamma'}\, \GF(z,z') D^{B'}D^{A'}\sigma_{A'B'}
	.
	\label{Q6}
\end{align}
In the second term on the \rhs, let us partial integrate the two covariant derivatives on $\sigma_{A'B'}$ to $\GF$.
This gives rise to two boundary terms, but one can use \eqref{Green} to show that they vanish, see appendix \ref{app:bdyGreen} for details,
\begin{align}
	\nQE
	&=
		\frac{1}{8\pi M}\int_\H d^2z\, (\p_z\p_\zb f) \int d^2z'\sqrt{\gamma'}\, (D^{A'}D^{B'}\GF(z,z')) \sigma_{A'B'}
	.
\end{align}
First, let us compute the Dirac bracket $\{\nQE,\iQE[g]\}_D$.
This is zero, since it is proportional to the expression
\begin{align}
	&\left\{
		\int_\H dv\,d^2z (\p_{z}\p_{\zb}f) \int d^2z'\sqrt{\gamma'} (D^{A'}D^{B'}\GF(z,z'))\sigma_{A'B'}
		,
		\int_\H dv\, d^2z''\sqrt{\gamma''}(D^{E''}D^{C''} g) \sigma_{D''C''}
	\right\}_D
	\nonumber
\end{align}
that vanishes.
Next, we compute $\{\nQE,\iQM[g]\}_D$.
It is proportional to the quantity
\begin{align}
	&
	\left\{
		\int_\H dv\,d^2z (\p_{z}\p_{\zb}f) \int d^2z'\sqrt{\gamma'} (D^{A'}D^{B'}\GF(z,z'))\sigma_{A'B'}
		,
		\int_\s d^2z''\sqrt{\gamma''}(D^{E''}D^{C''} g) \ep_{E''}{}^{D''} h_{D''C''}
	\right\}_D
	\nonumber\\ &=
		32\pi M^2
		\int_\s d^2z\, (\p_{z}\p_{\zb}f)
		\int d^2z'\sqrt{\gamma'} (D^{A'}D^{B'}\GF(z,z'))(D^{E'}D^{C'} g) \ep_{E'}{}^{D'}
		\gamma_{A'B'D'C'}
	,
\end{align}
where we have used \eqref{bracket},
	with $\gamma_{ABCD}=\gamma_{AC}\gamma_{BD}+\gamma_{AD}\gamma_{BC}-\gamma_{AB}\gamma_{CD}$.
Partial integrating the two covariant derivatives on $\GF$ to $g$ while noting that $D_A \ep_{BC} = 0$ and $D_A \gamma_{BCDE} = 0$, we obtain
\begin{align}
	&
	\left\{
		\int_\H dv\, d^2z (\p_{z}\p_{\zb}f) \int d^2z'\sqrt{\gamma'} (D^{A'}D^{B'}\GF(z,z'))\sigma_{A'B'}
		,
		\int_\s d^2z''\sqrt{\gamma''}(D^{E''}D^{C''} g) \ep_{E''}{}^{D''}h_{D''C''}
	\right\}_D
	\nonumber\\ &=
		32\pi M^2
		\int d^2z\, (\p_z\p_{\zb}f) \int d^2z'\sqrt{\gamma'} \GF(z,z')(D^{B'}D^{A'}D^{E'}D^{C'} g) \ep_{E'}{}^{D'}
		\gamma_{A'B'D'C'}
	\nonumber\\ &=
		64\pi iM^2
		\int d^2z\, (\p_z\p_{\zb}f) \int d^2z'\sqrt{\gamma'} \GF(z,z')
		\left(
			D^{\zb'}D^{\zb'}D^{z'}D^{z'} g
			-D^{z'}D^{z'}D^{\zb'}D^{\zb'} g
		\right)
			(\gamma_{z'\zb'})^2
	\nonumber\\ &=
		64\pi iM^2
		\int d^2z\, (\p_z\p_{\zb}f) \int d^2z'\sqrt{\gamma'} \GF(z,z')
		(\gamma^{z'\zb'})^2
		[D_{z'}^2,D_{\zb'}^2]g
	.	
\end{align}
The boundary term arising from the partial integration is similar to that discussed in appendix \ref{app:bdyGreen} and vanish for the same reason.\footnote{
The boundary term arising from the partial integration is proportional to the expression
\begin{align}
	&\int d^2z'\sqrt{\gamma'}
		\bigg[
			D^{A'}\left(D^{B'}\GF(z,z')D^{E'}D^{C'} g\right)
			-D^{B'}\left( \GF(z,z')D^{A'}D^{E'}D^{C'} g \right)
		\bigg]
		\ep_{E'}{}^{D'}\gamma_{A'B'D'C'}
	\nonumber\\&=
		2i\oint_z dz'
			\gamma^{z'\zb'}
			\left(\left (D_z^2 g\right )\p_{\zb'}\GF(z,z')-\GF(z,z')D_{\zb'}D_z^2 g\right)
		+ 2i\oint_z d\zb'
			\gamma^{z'\zb'}
			\left(\left (D_{\zb'}^2 g\right )\p_{z'}\GF(z,z') - \GF(z,z')D_{z'}D_{\zb'^2} g\right)
		.
	\nonumber
\end{align}
It is shown in appendix \ref{app:green} that $\GF\sim \frac{1}{4}\log|z-z'|^2$ as $z\to z'$, so the above expression vanishes due to lack of appropriate poles.
}
In the second equation, we have used the fact that the only non-vanishing components of $\ep_A{}^B$ and $\gamma_{ABCD}$ are
	$\ep_z{}^z = -\ep_\zb{}^\zb = i$ and $\gamma_{zz\zb\zb}=\gamma_{\zb\zb z z}=\frac{8}{(1+z\zb)^2}=2\gzz^2$ respectively.
One can readily check that $[D_z^2,D_\zb^2] g = 0$.

We conclude that $\nQE$ has zero bracket with both charges,
\begin{align}
	\{\nQE,\iQE[g]\}_D&=0
	,\qquad
	\{\nQE,\iQM[g]\}_D=0,
\end{align}
and therefore we do not be concerned about this term when computing Dirac brackets.

\subsection{Green's function for $D^2-1$}\label{app:green}

In this section, we present a derivation of the green's function for the negative-definite operator $D^2-1$ on the unit sphere using standard textbook techniques.
As operators of this form are of interest in various areas of physics, their Green's functions can be found in many places in the literature,
	see for example \cite{Szmytkowski2006} and references therein.

The Green's function $\GF(\Omega,\Omega')$ for $D^2-1$ is a solution to the equation
\begin{align}
	(D^2-1)\GF(\Omega,\Omega') = \delta(\Omega-\Omega') \equiv \frac{1}{\sin\t}\delta(\t-\t')\delta(\phi-\phi')
	\label{diff0}
	,
\end{align}
where $\Omega$ and $\Omega'$ represent points on the unit sphere, and the differential operator acts on $\Omega$.
Due to spherical symmetry, the Green's function will only depend on the geodesic distance between $\Omega$ and $\Omega'$.
Without any loss of generality, we can assign the coordinates on the sphere such that $\Omega'$ sits at the north pole.
Then, the geodesic distance between $\Omega$ and $\Omega'$ is given by $\t$.
By spherical symmetry, this solution must be the same as when $\Omega'$ is not necessarily at the north pole but instead $\phi=\phi'$,
	in which case the geodesic distance is $|\t-\t'|$.
Thus, we will solve the following equation first,
\begin{align}
	(D^2-1)\GF(|\t-\t'|) = \frac{1}{2\pi \sin\t}\delta(\t-\t')
	\label{diff1}
	,
\end{align}
and restore the $\phi$-dependence later.
The operator $D^2$ in spherical coordinates reads
\begin{align}
	D^2 = \frac{1}{\sin\t}\frac{\p}{\p \t}\sin\t\frac{\p}{\p \t} + \frac{1}{\sin^2\t}\frac{\p^2}{\p\phi^2},
\end{align}
so by changing variables to $t=\cos\t$, we can write \eqref{diff1} as
\begin{align}
	\left(\frac{d}{dt}(1-t^2)\frac{d}{dt} - 1\right)\GF(t,t') = \frac{1}{2\pi}\delta(t-t').
	\label{diff2}
\end{align}
We can obtain the Green's function $\GF$ by solving this equation for $t< t'$ and $t>t'$, and then stitching the two solutions together at $t=t'$.

The differential equation \eqref{diff2} states that a second-order differential operator acting on $\GF$ yields a delta function.
This implies that $\GF$ is continuous at $t=t'$; otherwise the discontinuity can locally be written in terms of the Heaviside step function,
	and $\frac{d^2}{dt^2}$ acting on it will yield a derivative of the delta function, which is not present in \eqref{diff2}.
So, we have
\begin{align}
	\lim_{\ep\to 0^+}\GF(t'-\ep,t') = \lim_{\ep\to 0^+} \GF(t'+\ep,t')
	\label{s1}
	.
\end{align}
On the other hand, $\frac{d\GF}{dt}$ is discontinuous, which can be seen by integrating \eqref{diff2} around an infinitesimal region around $t=t'$,
\begin{align}
	\lim_{\ep\to 0^+}(1-t'^2)\left(\left.\frac{d\GF}{dt}\right|_{t=t'+\ep} - \left.\frac{d\GF}{dt}\right|_{t=t'-\ep}\right) = \frac{1}{2\pi}.
	\label{s2}
\end{align}

With the stitching conditions \eqref{s1} and \eqref{s2} in mind, let us solve \eqref{diff2} for $t\neq t'$.
Equation \eqref{diff2} for $t\neq t'$ takes the form of a Legendre equation,
\begin{align}
	\left(
		\frac{d}{dt}(1-t^2)\frac{d}{dt}
		+ \lambda(\lambda+1)
	\right) \GF(t,t')
	= 0,
	\label{diff3}
\end{align}
with $\lambda = \frac{-1\pm i\sqrt 3}{2}$ (such that $\lambda(\lambda+1)=-1$).
Being a second-order ordinary differential equation, this has two linearly independent solutions,
	the Legendre functions $P_\lambda(t)$ and $Q_\lambda(t)$ of the first and second kind.
When $\lambda=n$ where $n$ is an integer, $P_n(t)$ is a Legendre polynomial. 
Legendre polynomials have a definite parity, so for instance $P_n(t)$ and $P_n(-t)=(-1)^nP_n(t)$ are not linearly independent.
However, for non-integer $\lambda$, $P_\lambda(t)$ is linearly independent to $P_\lambda(-t)$,
	(eqns$.$ 8.2.3 and 8.3.1 of \cite{Abramowitz1974})
\begin{align}
	P_\lambda(-t) = \cos(\lambda \pi)P_\lambda(t) - \frac{2}{\pi}\sin(\pi\lambda)Q_\lambda(t).
	\label{PQ}
\end{align}
This relation implies that for non-integer $\lambda$, we can use $P_\lambda(t)$ and $P_\lambda(-t)$ (instead of the standard pair $P_\lambda(t)$ and $Q_\lambda(t)$)
	as a basis of solutions to \eqref{diff3}.
Thus, we can write
\begin{align}
	\GF(t,t') =
	\begin{cases}
		a_1 P_\lambda(t) + a_2 P_\lambda(-t) &\qquad\text{for $t<t'$,}\\
		b_1 P_\lambda(t) + b_2 P_\lambda(-t) &\qquad\text{for $t>t'$,}
	\end{cases}
\end{align}
where $a_1$, $a_2$, $b_1$ and $b_2$ are functions of $t'$ only.
We demand that the Green's function $\GF(t,t')$ is well-defined everywhere but $t=t'$.
Taking note that $P_\lambda(1)=1$ and $P_\lambda(-1)=\infty$, one can see that this fixes $a_1 = b_2=0$,
\begin{align}
	\GF(t,t') =
	\begin{cases}
		a_2 P_\lambda(-t) &\qquad\text{for $t<t'$,}\\
		b_1 P_\lambda(t) &\qquad\text{for $t>t'$.}
	\end{cases}
	\label{G2}
\end{align}
The remaining coefficients $a_2$ and $b_1$ are fixed by the stitching conditions \eqref{s1} and \eqref{s2}, which read
\begin{align}
	a_2 P_\lambda(-t') &= b_1 P_\lambda(t'),
	\\
	b_1 P_\lambda'(t') + a_2 P'_\lambda(-t') &= \frac{1}{2\pi(1-t'^2)}.
\end{align}
These can equivalently be written as
\begin{align}
	\begin{pmatrix}
		P_\lambda(-t')&-P_\lambda(t')\\
		P'_\lambda(-t')&P_\lambda'(t')
	\end{pmatrix}
	\begin{pmatrix}
		a_2\\b_1
	\end{pmatrix}
	=
	\begin{pmatrix}
		0\\
		\frac{1}{2\pi(1-t'^2)}
	\end{pmatrix}
	.
\end{align}
Solving for $a_2$ and $b_1$, we obtain
\begin{align}
	\begin{pmatrix}
		a_2\\b_1
	\end{pmatrix}
	&=
		\frac{1}{(P_\lambda(-t')P_\lambda'(t')+P_\lambda(t')P'_\lambda(-t'))}
		\begin{pmatrix}
			P_\lambda'(t')&P_\lambda(t')\\
			-P'_\lambda(-t')&P_\lambda(-t')
		\end{pmatrix}
		\begin{pmatrix}
			0\\
			\frac{1}{2\pi(1-t'^2)}
		\end{pmatrix}
	\nonumber\\ &=
		\frac{-1}{2\pi(1-t'^2) \mW\{P_\lambda(t),P_\lambda(-t)\}|_{t=t'}}
		\begin{pmatrix}
			P_\lambda(t')\\
			P_\lambda(-t')
		\end{pmatrix}
	,
	\label{ab}
\end{align}
where $\mW\{\cdot,\cdot\}|_{t=t'}$ is the Wronskian,
\begin{align}
	\mW\{P_\lambda(t),P_\lambda(-t)\}
	&=
	\begin{vmatrix}
		P_\lambda(t) & P_\lambda(-t) \\
		\frac{d}{dt}P_\lambda(t) & \frac{d}{dt}P_\lambda(-t)
	\end{vmatrix}
	=
	\begin{vmatrix}
		P_\lambda(t) & P_\lambda(-t) \\
		P'_\lambda(t) & -P'_\lambda(-t)
	\end{vmatrix}
	\nonumber\\&=
		-\left(
			P_\lambda(t)P_\lambda'(-t)
			+P_\lambda(-t)P'_\lambda(t)
		\right)
		,
\end{align}
evaluated at $t=t'$.
To compute the Wronskian of $P_\lambda(t)$ and $P_\lambda(-t)$, we first note that the Wronskian of $P_\lambda(t)$ and $Q_\lambda(t)$ is (eqn$.$ 8.1.9 of \cite{Abramowitz1974})
\begin{align}
	\mW\{P_\lambda(t),Q_\lambda(t)\} = \frac{1}{1-t^2}.
\end{align}
Then, we use the relation \eqref{PQ} to obtain
\begin{align}
	\mW\{P_\lambda(t),P_\lambda(-t)\}
	&= \cos(\lambda\pi)\mW\{P_\lambda(t),P_\lambda(t)\} - \frac{2}{\pi}\sin(\pi\lambda)\mW\{P_\lambda(t),Q_\lambda(t)\}
	\nonumber\\ &= \frac{-2\sin(\pi \lambda)}{\pi (1-t^2)},
\end{align}
since $\mW\{P_\lambda(t), P_\lambda(t)\}=0$.
This with \eqref{ab} implies that $a_2$ and $b_1$ are
\begin{align}
	\begin{pmatrix}
		a_2\\b_1
	\end{pmatrix}
	&=
		\frac{1}{4\sin(\pi\lambda)}
		\begin{pmatrix}
			P_\lambda(t')\\
			P_\lambda(-t')
		\end{pmatrix}
	.
\end{align}
Plugging these into \eqref{G2}, we obtain the Green's function
\begin{align}
	\GF(t,t') =
	\frac{1}{4\sin(\pi\lambda)}
	\begin{cases}
		P_\lambda(t')P_\lambda(-t) &\qquad\text{for $t<t'$,}\\
		P_\lambda(-t')P_\lambda(t) &\qquad\text{for $t>t'$.}
	\end{cases}
\end{align}
Putting $\Omega'$ back at the north pole (and hence $\t'=0$ and $t'=1$) and recalling that $\lambda=\frac{-1+i\sqrt 3}{2}$, we obtain
\begin{align}
	\GF(\t) = \frac{1}{4\sin(\pi\lambda)} P_{\frac{-1+i\sqrt 3}{2}}(-\cos\t).
\end{align}
So, this is the Green's function when $\Omega'$ is the north pole.
For a generic point $\Omega'$ on the sphere, spherical symmetry demands that $\GF$ only depend on the geodesic distance $\gamma$ between $\Omega$ and $\Omega'$, which is given as
\begin{align}
	\cos\gamma = \cos\t \cos\t' + \sin\t\sin\t'\cos(\phi-\phi'),
	\label{angle}
\end{align}
and we have
\begin{align}
	\GF(\Omega,\Omega') = \frac{1}{4\sin(\pi\lambda)} P_{\frac{-1+i\sqrt 3}{2}}(-\cos\gamma),
\end{align}
as a solution to the equation \eqref{diff0}.
We note that it does not matter which of the two orders $\lambda=\frac{-1\pm i\sqrt 3}{2}$ we choose,
	since $P_{\lambda}(t) = P_{\lambda^*}(t)$; we have just chosen plus sign for definiteness.

\subsection{Treatment of boundary term} \label{app:bdyGreen}

In this section, we show that the boundary terms arising from partial integrating the r.h.s$.$ of \eqref{Q6} vanish.

One can see that this partial integration involves
\begin{align}
	&\int d^2z'\sqrt{\gamma'} \GF(z,z') D^{B'}D^{A'}\sigma_{A'B'}
	\nonumber\\ &=
		\int d^2z'\sqrt{\gamma'} D^{B'}\left(\GF(z,z') D^{A'}\sigma_{A'B'}\right)
		- \int d^2z'\sqrt{\gamma'} D^{A'}\left(\sigma_{A'B'}D^{B'}\GF(z,z')\right)
		\nonumber\\&\quad
		+ \int d^2z'\sqrt{\gamma'} \left(D^{A'}D^{B'}\GF(z,z')\right)\sigma_{A'B'}
	,
\end{align}
so the boundary term arising from this procedure is proportional to the quantity
\begin{align}
	&\int d^2z'\sqrt{\gamma'} D^{B'}\left(\GF(z,z') D^{A'}\sigma_{A'B'}\right)
	- \int d^2z'\sqrt{\gamma'} D^{A'}\left(\sigma_{A'B'}D^{B'}\GF(z,z')\right)
	\nonumber\\ &=
		- i\oint_z dz'\gamma^{z'\zb'} \left(\GF(z,z') \p_{\zb'}\sigma_{z'z'}-\sigma_{z'z'}\p_{\zb'}\GF(z,z')\right)
		\nonumber\\&\quad
		+ i\oint_z d\zb'\gamma^{z'\zb'} \left(\GF(z,z') \p_{z'}\sigma_{\zb'\zb'}-\sigma_{\zb'\zb'}\p_{z'}\GF(z,z')\right)
	,
	\label{boundary1}
\end{align}
where we have used Stokes' theorem.
This vanishes if (a) $\GF$ and $\p_{\zb'}\GF$ do not have $z'$-poles at $z'=z$ and (b) $\GF$ and $\p_{z'}\GF$ do not have $\zb'$-poles at $z'=z$.

To show that both (a) and (b) are true, we start from the Green's function $\GF(z,z')$ given in \eqref{Green}.
For the moment, let us put $z',\zb'=0$ (the north pole) and restore them later.
This gives
\begin{align}
	\GF(z,0)
	&= \frac{1}{4\sin(\lambda\pi)}P_\lambda \left(\frac{z\zb-1}{z\zb+1}\right)
	\label{GB}
	.
\end{align}
Only the asymptotic behavior of $\GF(z,0)$ near $z,\zb=0$ is relevant for the boundary contribution \eqref{boundary1},
	and for this we need the asymptotic behavior of $P_\lambda(t)$ near $t=-1$.
This can be derived via the asymptotic behaviors of $P_\lambda(t)$ and $Q_\lambda(t)$ near $t=1$, which read \cite{DLMF}
\begin{align}
	P_\lambda(t) &\sim 1
	,\qquad
	Q_\lambda(t) \sim \frac{1}{2}\ln\left(\frac{2}{1-t}\right),\qquad \text{as $t\to 1$,}
\end{align}
and using the relation \eqref{PQ}, which yields
\begin{align}
	P_{\lambda}(t) &\sim \frac{1}{\pi}\sin(\pi\lambda)\ln\left(1+t\right),\qquad\text{as $t\to -1$.}
	\label{Pasymp-1}
\end{align}
Applying this to the Green's function \eqref{GB} with $t=(z\zb-1)/(z\zb+1)$, we obtain
\begin{align}
	\GF(z,0) \sim \frac{1}{4}\ln\left (z\zb\right ),\qquad \text{as $z,\zb\to 0$.}
\end{align}
Restoring the reference point $z'$, the asymptotic form of the Green's function near $z=z'$ is\footnote{
	One can also derive this without putting $z'=0$ in the first place.
	To do so, one notes that $\cos\gamma$ in \eqref{angle} for generic $z$ and $z'$ can be obtained by taking the dot product of two vectors of the form \eqref{vector},
		and that it satisfies
	\begin{align}
		1 - \cos\gamma = 1 - \V n_z\cdot \V n_{z'}
		= \frac{2(z'-z)(\zb'-\zb)}{(1+z\zb)(1+z'\zb')}.
	\end{align}
	Then, taking $z=z'+r e^{i\phi}$ and expanding around $r=0$ leads to
	\begin{align}
		1 - \cos\gamma = \frac{2r^2}{(1+z'\zb')^2} + O(r^3),
	\end{align}
	which plugged into \eqref{Pasymp-1} for $P_\lambda(-\cos\gamma)$ and then into \eqref{Green} leads to $\GF(z,z')\sim \frac{1}{4}\ln r^2 = \frac{1}{4}\ln(z-z')(\zb-\zb')$ for $r\to 0$,
		in agreement with \eqref{asympG}.
}
\begin{align}
	\GF(z,z') \sim \frac{1}{4}\ln|z-z'|^2,\qquad \text{as $(z,\zb)\to (z',\zb')$.}
	\label{asympG}
\end{align}
One immediately sees that $\GF$ has a logarithmic singularity at $z=z'$ and therefore has no poles there.
Also, $\p_{z'}\GF = \frac{1}{4(z'-z)}$ has no $\zb'$-pole at $z'=z$, and $\p_{\zb'}\GF = \frac{1}{4(\zb'-\zb)}$ has no $z'$-pole at $z=z'$.
Therefore, the boundary term \eqref{boundary1} receives no residues and vanishes.

\bibliographystyle{jhep}
\bibliography{references}

\end{document}